\documentclass[draft,onecolumn,12pt]{IEEEtran}
\usepackage{enumerate,cite,amsmath,amsfonts,amssymb,subfigure,epsfig,times,graphicx,psfrag}
\usepackage{float}
\usepackage{verbatim,color}
\psfull
\newcommand{\mbf}{\boldmath}
\newcommand{\smbf}{\small \boldmath}

\def\ma{\text {max}}
\def\mi{\text {min}}

\def\sbQ{{\mbox {\smbf $Q$}}}
\def\bgamma{{\mblx {\mbf $\gamma$}}}
\def\b0{{\mbox {\mbf $0$}}}
\def\bA{{\mbox {\mbf $A$}}}
\def\ba{{\mbox {\mbf $a$}}}

\def\bh{{\mbox {\mbf $h$}}}

\def\bs{{\mbox {\mbf $s$}}}
\def\bu{{\mbox {\mbf $u$}}}

\def\bx{{\mbox {\mbf $x$}}}
\def\bv{{\mbox {\mbf $v$}}}
\def\by{{\mbox {\mbf $y$}}}
\def\bz{{\mbox {\mbf $z$}}}

\def\bE{{\mbox {\mbf $E$}}}
\def\bF{{\mbox {\mbf $F$}}}
\def\bH{{\mbox {\mbf $H$}}}
\def\bS{{\mbox {\mbf $S$}}}
\def\bU{{\mbox {\mbf $U$}}}
\def\bI{{\mbox {\mbf $I$}}}
\def\bR{{\mbox {\mbf $R$}}}
\def\bP{{\mbox {\mbf $P$}}}
\def\bQ{{\mbox {\mbf $Q$}}}
\def\bV{{\mbox {\mbf $V$}}}

\def\bG{{\mbox {\mbf $G$}}}

\def\bb0{{\mathbf{0}}}

\def\bgamma{\boldsymbol{\gamma}}

\def\bgamma{\mbox{\boldmath $\gamma$}}

\def\bgamma{\mbox{\boldmath $\gamma$}}
\def\b_beta{\mbox{\boldmath $\beta$}}
\def\hb_beta{\mbox{\boldmath $\hat \beta$}}

\newtheorem{Lemma}{Lemma}
\newtheorem{Corollary}{Corollary}
\newtheorem{example}{Example}
\newtheorem{definition}{Definition}
\newtheorem{Problem}{Problem}
\newtheorem{Remark}{Remark}
\newtheorem{Propo}{\it {Proposition}}
\begin{document}
\vfill
\title{Weighted Sum Rate Optimization for Cognitive Radio MIMO Broadcast Channels\\[.3cm]}
\author{Lan Zhang$^{\dagger}$, Yan Xin$^{\dagger *}$, and Ying-Chang
Liang$^{\ddagger}$\\[1.25cm]
%\thanks{
%$^*$ The work is supported by the National University of Singapore
%(NUS) under start-up grants R-263-000-314-101 and R-263-000-314-112
%and by a NUS Research Scholarship.}
\thanks{$^{\dag}$The authors are with the Department of Electrical and
Computer Engineering, National University of Singapore, Singapore
118622 (email: zhanglan@nus.edu.sg; elexy@nus.edu.sg).  The
corresponding author of the paper is Dr. Yan Xin (Tel. no: (65)
6516-5513 and Fax no. (65) 6779-1103).}
\thanks{${\ddag}$The author is with Institute of Infocomm Research, 21 Heng Mui Keng
Terrace, Singapore 119613 (email: ycliang@i2r.a-star.edu.sg).}
\thanks{Part of this work will be presented at IEEE International Conference on Communications, Beijing,
China, May 2008.}
\begin{abstract}
In this paper, we consider a cognitive radio (CR) network in which
the unlicensed (secondary) users are allowed to concurrently access
the spectrum allocated to the licensed (primary) users provided that
their interference to the primary users (PUs) satisfies certain
constraints. We study a weighted sum rate maximization problem for
the secondary user (SU) multiple input multiple output (MIMO)
broadcast channel (BC), in which the SUs have not only the sum power
constraint but also interference constraints. We first transform
this multi-constraint maximization problem into its equivalent form,
which involves a single constraint with multiple auxiliary
variables. Fixing these multiple auxiliary variables, we propose a
duality result for the equivalent problem. Our duality result can
solve the optimization problem for MIMO-BC with multiple linear
constraints, and thus can be viewed as an extension of the
conventional results, which rely crucially on a single sum power
constraint. Furthermore, we develop an efficient sub-gradient based
iterative algorithm to solve the equivalent problem and show that
the developed algorithm converges to a globally optimal solution.
Simulation results are further provided to corroborate the
effectiveness of the proposed algorithm.
\end{abstract}
\vspace*{-0.2cm}
\begin{keywords}\vspace*{-0.2cm}
Beamforming, broadcast channel, cognitive radio, MIMO, power
allocation, sum rate maximization.
\end{keywords}
\vspace*{-0.5cm} {\bf \footnotesize {EDICS}} \\ {\footnotesize
SPC-APPL: Applications involving signal processing for
communications \\[-0.25cm] MSP-MULT: MIMO multi-user and multi-access
schemes}} {}
\renewcommand{\thepage}{} \maketitle
%%%%%%%%%%%%%%%%%%%%%%%%%%%%%%%%%%%%%%%%%%%%%%%%%%%%%%%%%%%%%%%%%%%%%%%%\markboth{IEEE TRANSACTIONS ON SIGNAL PROCESSING (SUBMITTED)}
\newpage
\markboth{}{}
\renewcommand{\thepage}{} \maketitle \pagenumbering{arabic}
%\begin{abstract}
%%\setlength{\baselineskip}{1.2\baselineskip}
%In this paper, we consider a cognitive radio (CR) network in which
%the unlicensed (secondary) users are allowed to concurrently access
%the spectrum allocated to the licensed (primary) users provided that
%their interference to the primary users (PUs) satisfies certain
%constraints. We study a weighted sum rate maximization problem for
%the secondary user (SU) multiple input multiple output (MIMO)
%broadcast channel (BC), in which the SUs have not only the sum power
%constraint but also interference constraints. We first transform
%this multi-constraint maximization problem into its equivalent form,
%which involves a single constraint with multiple auxiliary
%variables. Fixing these multiple auxiliary variables, we establish a
%duality result for the equivalent problem. Our duality result can be
%viewed as an extension of the previously known results, which rely
%crucially on either a sum power constraint or per-antenna power
%constraints. Furthermore, we develop an efficient sub-gradient based
%iterative algorithm to solve the equivalent problem and show that
%the developed algorithm converges to a globally optimal solution.
%Simulation results are further provided to corroborate the
%effectiveness of the proposed algorithm.
%\end{abstract}
%%\vspace*{-0.2cm}
%\begin{keywords}
%Beamforming, broadcast channel, cognitive radio network, MIMO, power
%allocation, sum rate maximization.
%\end{keywords}

\section{Introduction}

Cognitive radio (CR), as a promising technology to advocate
efficient use of radio spectrum, has been a topic of increasing
research interest in recent years
\cite{FCC2003,Mitola1999,SimonKaykin:review,Xing_IEEE_TMC_2007,Gastpar_IEEE_TIT_2007,Ghasemi_Sousa_IEEE_TWC_2007,Liang:tradeoff08}.
CR allows an unlicensed (secondary) user to {\it opportunistically}
or {\it concurrently} access the spectrum initially allocated to the
licensed (primary) users provided that certain prescribed
constraints are satisfied, thus having a potential to improve
spectral utilization efficiency. In this paper, we study a weighted
sum rate maximization problem for the secondary user (SU) multiple
input multiple output (MIMO) broadcast channel (BC) in a concurrent
CR network, in which the SUs have not only the sum power constraint
but also interference constraints.

\subsection{System Model and Problem Formulation}

With reference to Fig. \ref{fig:sysmodel}, we consider the $K$-SU
MIMO-BC with $N_t$ transmit antennas and $N_r$ receive antennas in a
CR network, where the $K$ SUs share the same spectrum with a single
primary user (PU) equipped with one transmitter and one
receiver\footnote{Expect for explicitly stated, we restrict our
attention to a single PU case in the rest of this paper for
convenience of description. The results derived for the single PU
case can be readily extended to the multiple PU case, which is
discussed in Remark \ref{remark:multiPUs}.}. The transmit-receive
signal model from the BS to the $i$th SU denoted by SU$_i$, for
$i=1,\ldots, K$, can be expressed as
\begin{equation}\label{eq:model}
\by_i = \bH_i\bx + \bz_i,
\end{equation}
where $\by_i$ is the $N_r\times 1$ received signal vector, $\bH_i$
is the $N_r\times N_{t}$ channel matrix from the BS to the SU$_i$,
$\bx$ is the $N_{t}\times 1$ transmitted signal vector, and $\bz_i$
is the $N_r\times 1$ Gaussian noise vector with entries being
independent identically distributed random variables (RVs) with mean
zero and variance $\sigma^2$. Consider $\bh_{o}$ as the $N_t\times
1$ channel gain vector between the transmitters of the BS and the
PU. We further assume that $\bH_i$ for $i=1,\ldots,K$, and $\bh_{o}$
remain constant during a transmission block and change independently
from block to block, and $\bH_i$ for $i=1,\ldots,K$, and $\bh_{o}$
are perfectly known to the BS and SU$_i$. This requires that the SUs
can ``cognitively'' obtain the information of its neighboring
environment. In practice, certain cooperation in terms of parameter
feedback between the PU and the BS is needed. To achieve that, the
protocol for the SU network can be designed as follows: every frame
contains sensing sub-frame and data transmission sub-frame. During
the sensing sub-frame, BS can transmit training sequences to SUs as
well as to the PU so that the SUs can estimate the channel matrix
$\bH_i$, and the PU can measure the vector $\bh_o$. After that, this
information will be sent back to the BS via a feedback channel.

We next consider the weighted sum rate maximization problem for the
$K$-SU MIMO-BC in a CR network, simply called the CR MIMO-BC sum
rate maximization problem, which, mathematically, can be formulated
as
\begin{Problem}[Main Problem]\label{eq:objfun}
\begin{align}
\underset{{\{\sbQ_i^{\text{b}}\}_{i=1}^K:~\sbQ^{\text{b}}_i\succeq 0}}{\ma}~&\sum_{i=1}^{K}w_ir_i^{\text{b}}\label{eq:weightedsumrate}\\
\text{subject
to~}~&\sum_{i=1}^{K}\bh_{o}^{\dag}\bQ_i^{\text{b}}\bh_{o} \leq P_t,
~ \text{and}~\sum_{i=1}^{K}\text{tr}(\bQ_i^{\text{b}})\leq
P_u,\nonumber
\end{align}
\end{Problem}
where $r_i^{\text{b}}$ is the rate achieved by SU$_i$, $w_i$ is the
weight of SU$_i$, $\bQ_i^{\text{b}}$ denotes the $N_t\times N_t$
transmit signal covariance matrix for SU$_i$,
$\bQ_i^{\text{b}}\succeq 0$ denotes that $\bQ_i^{\text{b}}$ is a
semidefinite matrix, $P_t$ denotes the interference threshold of the
PU, and $P_u$ denotes the sum power constraint at the BS. In a
non-CR setting, similar weighted sum rate optimization problems for
the multiple input single output (MISO) BC and the MIMO-BC have been
studied in \cite{Caire06:MISOMACsumrate}\cite{Liujia:conjgrad},
respectively. The key difference is that in addition to the sum
power constraint, an interference constraint is applied to the SUs
in the CR MIMO-BC, i.e., the total received interference power
$\sum_{i=1}^{K}\bh_{o}^{\dag}\bQ_i^{\text{b}}\bh_{o}$ at the PU is
below the threshold $P_t$.

\begin{Remark}
It has long been observed that the optimal sum rate for MIMO BC with
a single sum power constraint is equal to the optimal sum rate of
the dual MIMO multiple access channel (MAC) with the same sum power
constraint
\cite{FarrokhiLiu1998}\cite{Viswanath2003:sumcapacity}\cite{Jindal03:sumcapacity}.
However, this {\it conventional BC-MAC duality} can only be applied
to the case with a single sum power constraint (even not applicable
to an arbitrary linear power constraint). Hence, the additional
interference power constraint in Problem \ref{eq:objfun} makes the
existing duality cannot be applied. The new duality result proposed
in this paper generalizes the previous results as special cases.
Moreover, it is worth to note that any boundary point of the
capacity regions of the MIMO-MAC and the MIMO-BC can be expressed as
a weighted sum rate for a certain choice of weights
\cite{Tomluo06:optimization} \cite{Tse98:Polymatriod}. Thus, by
varying the weights of the SUs in Problem \ref{eq:objfun}, the
entire capacity region of the CR MIMO-BC can be obtained.
\end{Remark}

\subsection{Related Work}

The present paper is motivated by the previous work on the
information-theoretic study of the MIMO-BC under a {\it non-CR}
setting. It has been shown in
\cite{Viswanath2003:sumcapacity}\cite{Jindal03:sumcapacity}\cite{Yuwei2004:broadcastthroughput}
that under a single sum power constraint, the sum-capacity of the
non-CR MIMO-BC can be achieved by the dirty paper coding (DPC)
scheme. Furthermore, the paper \cite{Shamai06:MIMOBCcapacity} shows
that the rate region achieved by the DPC scheme is indeed the
capacity region of such a channel. However, the power allocation and
beamforming strategies to achieve the capacity region have been not
considered in these papers. Moreover, it has been shown in
\cite{Jindal05:sumpowerMAC}\cite{Yuwei2006:sumcapacitycomputation}
that under the single sum power constraint, the {\it equally}
weighted sum rate maximization problem, simply called the sum rate
problem, for the MIMO-BC can be solved by solving its dual MIMO MAC
sum rate problem, which is also subject to a single sum power
constraint. In \cite{Jindal05:sumpowerMAC}, a cyclic coordinate
ascent algorithm was proposed to solve the dual MIMO-MAC problem
while in \cite{Yuwei2006:sumcapacitycomputation} this sum-power
constrained dual problem was decoupled into an individual-power
constrained problem, which can be solved by using an iterative
water-filling algorithm \cite{Yuwei2004:IWF_MAC}. Even though these
algorithms proposed in
\cite{Jindal05:sumpowerMAC}\cite{Yuwei2006:sumcapacitycomputation}
can solve the sum rate optimization problem for the non-CR MIMO-BC
via the MAC-BC duality, they are not applicable to the general
weighted sum rate problem. In \cite{Caire06:MISOMACsumrate}, a
generalized iterative water-filling was proposed to solve the
weighted sum rate problem for the MISO-BC where each user has a
single receive antenna. However, the proposed algorithm is not
applicable to the general MIMO-BC case. Furthermore, an efficient
algorithm was proposed to solve the MIMO-BC weighted sum rate
problem with a single sum power constraint in
\cite{Liujia:conjgrad}. These aforementioned results are based on
the conventional BC-MAC duality, which cannot be applied to solve
the weighted sum rate problem with multiple constraints (the case of
interest in this paper). Recently, the paper
\cite{Yuwei07:perantconst} investigated a different MIMO-BC weighted
sum rate maximization problem which is subject to per-antenna power
constraints instead of the single sum power constraint, and
established a new {\it minimax} duality which is different from the
conventional BC-MAC duality. A Newton's method based algorithm was
proposed to solve this minimax problem. In this paper, we consider a
more general case where the power is subject to multiple linear
constraints instead of the sum power constraint or per-antenna power
constraints, and propose a new BC-MAC duality result to extend the
conventional duality result so that it can solve the problem with
multiple arbitrary linear constraints. A Karush-Kuhn-Tucker (KKT)
condition based algorithm is developed to solve the problem.

\subsection{Contribution}

Throughout the paper, we consider the CR MIMO-BC weighted sum rate
maximization problem as defined in Problem \ref{eq:objfun}. As the
main contribution of this paper, our solution is summarized in the
following.

\begin{enumerate}

\item We prove that in the CR MIMO-BC, the multi-constraint
weighted sum rate maximization problem (Problem \ref{eq:objfun})
is equivalent to a single-constraint weighted sum rate
maximization problem with multiple auxiliary variables.

\item For the equivalent problem, we establish a duality between
the MIMO-BC and a dual MIMO-MAC when the multiple auxiliary
variables are fixed as constant. This duality is applicable to
MIMO-BC with arbitrary linear power constraint, and can be viewed as
an extension of the conventional MIMO MAC-BC duality result
\cite{FarrokhiLiu1998}\cite{Viswanath2003:sumcapacity}\cite{Jindal03:sumcapacity},
which is only valid for the problem with a single sum power
constraint.

\item For the weighted sum rate maximization problem of the dual MIMO MAC, the existing iterative water-filling based algorithm \cite{Jindal05:sumpowerMAC,Yuwei2006:sumcapacitycomputation} is not applicable. We propose a new primal dual method based iterative algorithm \cite{matrix_opt_book} to solve it. Furthermore, we propose a sub-gradient based iterative algorithm to solve the main problem of the paper, Problem
\ref{eq:objfun}, and show that the proposed algorithm converges to
the globally optimal solution.
\end{enumerate}

\subsection{Organization and Notation}

The rest of the paper is organized as follows. In Section
\ref{section:duality}, we transform the CR MIMO-BC weighted sum rate
maximization problem (Problem \ref{eq:objfun}) into its equivalent
form, and introduce a MAC-BC duality between a MIMO-BC and a dual
MIMO-MAC. Section \ref{section:weightsumrateMAC} presents an primal
dual method based iterative algorithm to solve the dual MIMO-MAC
weighted sum rate problem. In Section \ref{section:Mapping}, a
MAC-BC covariance matrix mapping algorithm is proposed. Section
\ref{section:overallalgorithm} presents the complete algorithm to
solve the CR MIMO-BC weighted sum rate maximization problem. Section
\ref{section:simulation} provides several simulation examples.
Finally, Section \ref{section:conclusion} concludes the paper.

The following notations are used in this paper. The boldface is used
to denote matrices and vectors, $(\cdot)^{\dag}$ and $(\cdot)^{T}$
denote the conjugate transpose and transpose, respectively;
${\bI}_{M}$ denotes an $M \times M$ identity matrix;
$\text{tr}(\cdot)$ denotes the trace of a matrix, and $[x]^+$
denotes $\max(x,0)$; $(\cdot)^{\text{b}}$ and $(\cdot)^{\text{m}}$
denote the quantities associated with a broadcast channel and a
multiple access channel, respectively; $E[\cdot]$ denotes the
expectation operator.

\section{Equivalence and Duality}\label{section:duality}

Evidently, the MIMO-BC weighted sum rate maximization problem under
either a non-CR or a CR setting is a non-convex optimization problem
and is difficult to solve directly. Under a single sum power
constraint, the weighted sum rate problem for MIMO BC can be
transformed to its dual MIMO MAC problem, which is convex and can be
solved in an efficient manner
\cite{Caire06:MISOMACsumrate}\cite{Liujia:conjgrad}. In the CR
setting, the problem (Problem \ref{eq:objfun}) has not only a sum
power constraint but also an interference constraint. The imposed
multiple constraints render difficulty to formulate an efficiently
solvable dual problem. To overcome the difficulty, we first
transform this multi-constrained weighted sum rate problem (Problem
\ref{eq:objfun}) into its equivalent problem which has a single
constraint with multiple auxiliary variables, and next develop a
duality between a MIMO-BC and a dual MIMO-MAC in the case where the
multiple auxiliary variables are fixed.

\subsection{An Equivalent MIMO-BC Weighted Sum Rate Problem}

In the following proposition, we present an equivalent form of
Problem \ref{eq:objfun} (see Appendix \ref{proposition:proof1} for
the proof).
\begin{Propo}\label{lemma:BC2BC}
Problem \ref{eq:objfun} shares the same optimal solution with
\begin{Problem}[Equivalent Problem]\label{eq:objfuntransform}
\begin{align}
&\underset{q_{t}\geq0,~q_{u}\geq 0}{\mi}\underset{{\{\sbQ_i^{\text{b}}\}_{i=1}^K:~\sbQ^{\text{b}}_i\succeq 0}}{\ma}~\sum_{i=1}^{K}w_ir_i^{\text{b}}\\
&\text{subject to~}~q_{t}  \big(
\sum_{i=1}^{K}\bh_{o}^{\dag}\bQ_i^{\text{b}}\bh_{o} -P_t \big)
+q_{u} \big( \sum_{i=1}^K\text{tr}(\bQ_i^{\text{b}})-P_u \big)
\leq 0,
\end{align}
\end{Problem}
where $q_{t}$ and $q_{u}$ are the auxiliary dual variables for the
respective interference constraint and sum power constraint.
\end{Propo}

It can be readily concluded from the proposition that the optimal
solution to Problem \ref{eq:objfuntransform} also satisfies
$\sum_{i=1}^{K}\bh_{o}^{\dag}\bQ_i^{\text{b}}\bh_{o} \leq P_t$ and
$\sum_{i=1}^{K}\text{tr}(\bQ_i^{\text{b}})\leq P_u$ simultaneously
since it is also the optimal solution to Problem \ref{eq:objfun}.
Finding an efficiently solvable dual problem for Problem
\ref{eq:objfuntransform} directly is still difficult. However, as we
show later, when $q_t$ and $q_u$ are fixed as constants, Problem
\ref{eq:objfuntransform} reduces to a simplified form, which we can
solve by applying the following duality result.

\subsection{CR MIMO BC-MAC Duality}\label{subsection:MACBCdual}

For fixed $q_t$ and $q_u$, Problem \ref{eq:objfuntransform}
reduces to the following form
\begin{Problem}[CR MIMO-BC]\label{Prob:fixq}
\begin{align}
\underset{{\{\sbQ_i^{\text{b}}\}_{i=1}^K:~\sbQ^{\text{b}}_i\succeq 0}}{\ma}~&\sum_{i=1}^{K}w_ir_i^{\text{b}}\label{eq:prob3obj}\\
\text{subject to~}~q_{t}
&\sum_{i=1}^{K}\bh_{o}^{\dag}\bQ_i^{\text{b}}\bh_{o}   +q_{u}
\sum_{i=1}^K\text{tr}(\bQ_i^{\text{b}}) \leq P,
\label{eq:transformBCcons}
\end{align}
\end{Problem}
where $P:=q_{t}P_{t}+q_{u}P_{u}$. Since $q_t$ and $q_u$ are fixed,
$P$ is a constant in Problem \ref{Prob:fixq}. The constraint
\eqref{eq:transformBCcons} is not a single sum power constraint, and
thus the duality result established in \cite{Jindal05:sumpowerMAC}
is not applicable to Problem \ref{Prob:fixq}. Therefore, we
formulate the following new dual MAC problem.
\begin{Propo}\label{proposition:dual}
The dual MAC problem of Problem \ref{Prob:fixq} is
\begin{Problem}[CR MIMO-MAC]\label{eq:objfunMACdual}
\begin{align}
\underset{{\{\sbQ_i^{\text{m}}\}_{i=1}^K:~\sbQ^{\text{m}}_i\succeq 0}}{\ma}~&\sum_{i=1}^{K}w_ir_i^{\text{m}}\label{eq:macobj}\\
\text{subject
to~}~&\sum_{i=1}^K\text{tr}(\bQ_i^{\text{m}})\sigma^2\leq
P,\label{eq:dualMACcons}
\end{align}
\end{Problem}
\end{Propo}
where $r_i^{\text{m}}$ is the rate achieved by the $i$th user of the
dual MAC, $\bQ_i^{\text{m}}$ is the transmit signal covariance
matrix of the $i$th user, and the noise covariance at the BS is
$q_{t}\bh_o\bh_o^H+q_{u}\bI_{N_t}$.

\begin{Remark}
According to Proposition \ref{proposition:dual},  for fixed $q_t$
and $q_u$, the optimal weighted sum rate of the dual MAC is equal to
the optimal weighted sum rate of the primal BC. From the formulation
perspective, this duality result is quite similar to the
conventional duality in
\cite{FarrokhiLiu1998}\cite{Viswanath2003:sumcapacity}\cite{Jindal03:sumcapacity}.
However, as shown in Fig. \ref{fig:sysmodelMAC}, one thing needs to
highlight is that the noise covariance matrix of the dual MAC is a
function of the auxiliary variable $q_t$ and $q_u$, instead of the
identity matrix \cite{Jindal03:sumcapacity}. This difference comes
from the constraint \eqref{eq:transformBCcons}, which is not a sum
power constraint as in \cite{Jindal03:sumcapacity}. Note that when
$q_t=0$, the duality result reduces to the conventional BC-MAC
duality in \cite{Jindal03:sumcapacity}.
\end{Remark}

As illustrated in Fig. \ref{fig:sysmodelMAC}, Proposition
\ref{proposition:dual} describes a weighted sum rate maximization
problem for a dual MIMO-MAC. To prove the proposition, we first
examine the relation between the {\it signal to interference plus
noise ratio} (SINR) regions of the MIMO-BC and the dual MIMO-MAC.
Based on this relation, we will show that the achievable rate
regions of the MIMO-BC and the dual MIMO-MAC are the same.

In the sequel, we first describe the definition of the $\text{SINR}$
for the MIMO-BC. It has been shown in \cite{Shamai06:MIMOBCcapacity}
that the DPC is a capacity achieving scheme. Each set of the
transmit covariance matrix determined by DPC scheme defines a set of
transmit and receive beamforming vectors, and each pair of these
transmit and receive beamforming vectors forms a data stream. In a
beamforming perspective, the BS transmitter have $N_t\times K$
beamformers, $\bu_{i,j}$, for $i=1,\cdots,K$, and $j=1,\cdots,N_t$.
Therefore, the transmit signal can be represented as
\begin{equation*}
\bx=\sum_{i=1}^{K}\sum_{j=1}^{N_t}x_{i,j}\bu_{i,j},
\end{equation*}
where $x_{i,j}$ is a scalar representing the data stream transmitted
in this beamformer, and $E[x_{i,j}^2]=p_{i,j}$ denotes the power
allocated to this beamformer. At SU$_i$, the receive beamformer
corresponding to $\bu_{i,j}$ is denoted by $\bv_{i,j}$. The transmit
beamformer $\bu_{i,j}$ and the power $p_{i,j}$ can be obtained via
the eigenvalue decomposition of $\bQ_i^{\text{b}}$, i.e.,
$\bQ_i^{\text{b}} = \bU_i^{\dag}\bP_i\bU_i$, where $\bU_i$ is a
unitary matrix, and $\bP_i$ is a diagonal matrix. The transmit
beamformer $\bu_{i,j}$ is the $j$th column of $\bU_i$, and $p_{i,j}$
is the $j$th diagonal entry of $\bP_i$. With these notations, we
express the $\text{SINR}_{i,j}^{\text{b}}$ as

\begin{equation}\label{eq:sinrbc}
\text{SINR}_{i,j}^{\text{b}}=\frac{p_{i,j}|\bu_{i,j}^{\dag}\bH_i^{\dag}\bv_{i,j}|^2}
{\sum_{k=i+1}^{K}\sum_{l=1}^{N_{r}}p_{k,l}|\bu_{k,l}^{\dag}\bH_i^{\dag}\bv_{i,j}|^2+
\sum_{l=j+1}^{N_{r}}p_{i,l}|\bu_{i,l}^{\dag}\bH_i^{\dag}\bv_{i,j}|^2+\sigma^2}.
\end{equation}
It can be observed from \eqref{eq:sinrbc} that the DPC scheme is
applied. This can be interpreted as follows. The signal from
SU$_{1}$ is first encoded with the signals from other SUs being
treated as interference. The signal from SU$_2$ is next encoded by
using the DPC scheme. Signals from the other SUs will be encoded
sequentially in a similar manner. For the data streams within
SU$_i$, the data stream 1 is also encoded first while the other data
streams are treated as the interference. The data stream 2 is
encoded next. In a similar manner, the other data streams will be
sequentially encoded. The encoding order is assumed to be arbitrary
at this moment, and the optimal encoding order of Problem
\ref{eq:objfuntransform} will be discussed in Section
\ref{section:weightsumrateMAC}.

To explore the relation of the SINR regions of the dual MAC and the
BC, we formulate a following optimization problem
\begin{align}
\begin{split}
\underset{{\{\sbQ_i^{\text{b}}\}_{i=1}^K:~\sbQ^{\text{b}}_i\succeq
0}}{\mi}~&q_{t}\sum_{i=1}^{K}\bh_{o}^{\dag}\bQ_i^{\text{b}}\bh_{o}
+q_{u}\sum_{i=1}^{K}\text{tr}(\bQ_i^{\text{b}})- P\\
\text{subject to~}~&\text{SINR}_{i,j}^{\text{b}}\geq
\gamma_{i,j},\label{probform:BCoriginal1}
%\label{eq:BCoriginal1cons}
\end{split}
\end{align}
where $\gamma_{i,j}$ denotes the SINR threshold of the $j$th data
stream within the SU$_i$ for the BC. Note that the objective
function in \eqref{probform:BCoriginal1} is a function of signal
covariance matrices and the constraints are SINR constraints for the
$K$-SU MIMO-BC.

It has been shown in \cite{Yuwei07:perantconst} and
\cite{Shamai2006:conicoptimazation} that the {\it non-convex} BC sum
power minimization problem under the SINR constraints can be solved
efficiently via its dual MAC problem, which is a convex optimization
problem. By following a similar line of thinking, the problem in
\eqref{probform:BCoriginal1} can be efficiently solved via its dual
MAC problem. Similar to the primal MIMO-BC, the dual MIMO-MAC
depicted in Fig. \ref{fig:sysmodelMAC} consists of $K$ users each
with $N_r$ transmit antennas, and one BS with $N_t$ receive
antennas. By transposing the channel matrix and interchanging the
input and output signals, we obtain the dual MIMO-MAC from the
primal MIMO-BC. For the covariance matrices $\bQ_i^{\text{m}}$ of
the dual MIMO-MAC, we apply the eigenvalue decomposition,
\begin{equation}\label{eq:eigendecV}
\bQ^{\text{m}}_i=\bV_i{\bf
\Lambda}_i\bV^{\dag}_i=\sum_{j=1}^{N_r}q_{i,j}\bv_{i,j}\bv_{i,j}^{\dag},
\end{equation}
where $\bv_{i,j}$ is the $j$th column of $\bV_i$, and $q_{i,j}$ is
the $j$th diagonal entry of ${\bf{\Lambda}}_i$. For user $i$,
$\bv_{i,j}$ is the transmit beamforming vector of the $j$th data
stream, the power allocated to the $j$th data stream equals
$q_{i,j}$, and the receive beamforming vector of the $j$th data
stream at the BS is $\bu_{i,j}$. The SINR of the dual MIMO-MAC is
given by
\begin{equation}\label{eq:sinrmac}
\text{SINR}_{i,j}^{\text{m}}=
\frac{q_{i,j}|\bu_{i,j}^{\dag}\bH_i^{\dag}\bv_{i,j}|^2}{\bu_{i,j}^{\dag}\left(\sum_{k=1}^{i-1}\sum_{l=1}^{N_{r}}q_{k,l}\bH_k^{\dag}\bv_{k,l}\bv_{k,l}^{\dag}\bH_k+
\sum_{l=1}^{j-1}q_{i,l}\bH_i^{\dag}\bv_{i,l}\bv_{i,l}^{\dag}\bH_i+\bR_w\right)\bu_{i,j}},
\end{equation}
where $\bR_w:=q_{t}\bR_o+q_{u}\bI_{N_t}$ is the noise covariance
matrix of the MIMO-MAC with $\bR_o:=\bh_o\bh_o^{\dag}$. In the dual
MIMO-MAC, $\bR_w$ depends on $q_{t}$ and $q_{u}$ defined in
\eqref{probform:BCoriginal1} whereas the noise covariance matrix in
the primal MIMO-BC is an identity matrix. It can be observed from
\eqref{eq:sinrmac} that the successive interference cancelation
(SIC) scheme is used in this dual MIMO-MAC, and the decoding order
is the reverse encoding order of the primal BC. The signal from
SU$_K$ is first decoded with the signals from other users being
treated as interference. After decoded at the BS, the signals from
SU$_K$ will be subtracted from the received signal. The signal from
SU$_{K-1}$ is next decoded, and so on. Again, the data streams
within a SU can be decoded in a sequential manner.

For the dual MIMO-MAC, we consider the following minimization
problem similar to the problem \eqref{probform:BCoriginal1}
\begin{align}\label{probform:MACequil}
\begin{split}
\underset{{\{\sbQ_i^{\text{m}}\}_{i=1}^K:~\sbQ^{\text{m}}_i\succeq 0}}{\mi}~&\sum_{i=1}^K\text{tr}(\bQ_i^{\text{m}})\sigma^2-P\\
\text{subject to~}~ &\text{SINR}_{i,j}^{\text{m}}\geq
\gamma_{i,j}.%\label{ineq:MACcons}
\end{split}
\end{align}
The following proposition describes the relation between the
problems \eqref{probform:BCoriginal1} and \eqref{probform:MACequil}.
\begin{Propo}\label{thm:Macduality}
For fixed $q_{t}$ and $q_{u}$, the MIMO-MAC problem
\eqref{probform:MACequil} is dual to the MIMO-BC problem
\eqref{probform:BCoriginal1}.
\end{Propo}
\begin{proof}
The constraints in \eqref{probform:BCoriginal1} can be rewritten
as
\begin{align}
\frac{p_{i,j}|\bu_{i,j}^{\dag}\bH_i^{\dag}\bv_{i,j}|^2}{\gamma_{i,j}}\!\geq\!
\sum_{k=i+1}^{K}\sum_{l=1}^{N_{r}}p_{k,l}|\bu_{k,l}^{\dag}\bH_i^{\dag}\bv_{i,j}|^2\!+\!\sum_{l=j+1}^{N_{r}}p_{i,l}|\bu_{i,l}^{\dag}\bH_i^{\dag}\bv_{i,j}|^2+\sigma^2.
\end{align}
Thus, the Lagrangian function of the problem
\eqref{probform:BCoriginal1} is
\begin{align}
&L_1(\bQ^{\text{b}}_1,\ldots,\bQ^{\text{b}}_K,\lambda_{i,j})
\notag \\
=&q_{t}\sum_{i=1}^K\bh_{o}^{\dag}\bQ_i^{\text{b}}\bh_{o}+q_{u}\sum_{i=1}^K\text{tr}(\bQ_i^{\text{b}})-P
-\sum_{i=1}^{K}\sum_{j=1}^{N_{r}}\lambda_{i,j}\Big(\frac{p_{i,j}|\bu_{i,j}^{\dag}\bH_i^{\dag}\bv_{i,j}|^2}{\gamma_{i,j}}\notag
\\
&-\sum_{k=i+1}^{K}\sum_{l=1}^{N_{r}}p_{k,l}|\bu_{k,l}^{\dag}\bH_i^{\dag}\bv_{i,j}|^2
-\sum_{l=j+1}^{N_{r}}p_{i,l}|\bu_{i,l}^{\dag}\bH_i^{\dag}\bv_{i,j}|^2-\sigma^2\Big)\label{eq:BCLagran1}
\end{align}
\begin{align}
=&\sum_{i=1}^{K}\sum_{j=1}^{N_{r}}\lambda_{i,j}\sigma^2-P-\sum_{i=1}^{K}\sum_{j=1}^{N_{r}}p_{i,j}\bu_{i,j}^{\dag}\Big(\frac{\lambda_{i,j}\bH_i^{\dag}\bv_{i,j}\bv_{i,j}^{\dag}\bH_i}{\gamma_{i,j}}\notag\\
&-\sum_{k=1}^{i-1}\sum_{l=1}^{N_{r}}\lambda_{k,l}\bH_k^{\dag}\bv_{k,l}\bv_{k,l}^{\dag}\bH_k-\sum_{l=1}^{j-1}\lambda_{i,l}\bH_i^{\dag}\bv_{i,l}\bv_{i,l}^{\dag}\bH_i-\bR_w\Big)\bu_{i,j},\label{eq:BCLagran}
\end{align}
where $\lambda_{i,j}$ is the Lagrangian multiplier. Eq.
\eqref{eq:BCLagran} is obtained by applying the eigenvalue
decomposition to $\bQ_i^{\text{b}}$ and rearranging the terms in
\eqref{eq:BCLagran1}. The optimal objective value of
\eqref{probform:BCoriginal1} is
\begin{equation}
\underset{\lambda_{i,j}}{\ma}\underset{\sbQ^{\text{b}}_1,\ldots,\sbQ^{\text{b}}_K}{\mi}L_1(\bQ^{\text{b}}_1,\ldots,\bQ^{\text{b}}_K,\lambda_{i,j}).\label{eq:objvalue}
\end{equation}

On the other hand, the Lagrangian function of the problem
\eqref{probform:MACequil} is
\begin{align}
L_2(\bQ^{\text{m}}_1,\ldots,\bQ^{\text{m}}_K,\delta_{i,j})=\sum_{i=1}^{K}\sum_{j=1}^{N_{r}}q_{i,j}\sigma^2-P-\sum_{i=1}^{K}\sum_{j=1}^{N_{r}}\delta_{i,j}\bu_{i,j}^{\dag}(\frac{q_{i,j}\bH_i^{\dag}\bv_{i,j}\bv_{i,j}^{\dag}\bH_i}{\gamma_{i,j}}\notag\\
-\sum_{k=1}^{i-1}\sum_{l=1}^{N_{r}}q_{k,l}\bH_k^{\dag}\bv_{k,l}\bv_{k,l}^{\dag}\bH_k-\sum_{l=1}^{j-1}q_{i,l}\bH_i^{\dag}\bv_{i,l}\bv_{i,l}^{\dag}\bH_i-\bR_w)\bu_{i,j},\label{eq:MACLagran}
\end{align}
where $\delta_{i,j}$ is the Lagrangian multiplier. Eq.
\eqref{eq:MACLagran} is also obtained by applying eigenvalue
decomposition to $\bQ_i^m$. The optimal objective value of
\eqref{probform:MACequil} is
\begin{equation}
\underset{\delta_{i,j}}{\ma}\underset{\sbQ^{\text{m}}_1,\ldots,\sbQ^{\text{m}}_K}{\mi}L_2(\bQ^{\text{m}}_1,\ldots,\bQ^{\text{m}}_K,\delta_{i,j}).\label{eq:objvalue1}
\end{equation}

Note that if we choose $q_{i,j}=\lambda_{i,j}$,
$\delta_{i,j}=p_{i,j}$, and the same beamforming vectors
$\bu_{i,j}$ and $\bv_{i,j}$ for both problems, \eqref{eq:BCLagran}
and \eqref{eq:MACLagran} become identical. This means that the
optimal solutions of \eqref{eq:objvalue} and \eqref{eq:objvalue1}
are the same.
\end{proof}

Proposition \ref{thm:Macduality} implies that under the  SINR
constraints, the problems \eqref{probform:BCoriginal1} and
\eqref{probform:MACequil} can achieve the same objective value,
which is a function of the transmit signal covariance matrices. On
the other hand, under the corresponding constraints on the signal
covariance matrix, the achievable SINR regions of the MIMO-BC and
its dual MIMO-MAC are the same. Mathematically, we define the
respective achievable SINR regions for the primal MIMO-BC and the
dual MIMO-MAC as follows.
\begin{definition}
A SINR vector
$\bgamma=(\gamma_{1,1},\ldots,\gamma_{1,N_t},\ldots,\gamma_{K,N_t})$
is said to be achievable for the primal BC if and only if there
exists a set of $\bQ_1^b,\ldots,\bQ_K^b$ such that
$q_{t}\sum_{i=1}^{K}\bh_{o}^{\dag}\bQ_i^{\text{b}}\bh_{o}
+q_{u}\sum_{i=1}^{K}\text{tr}(\bQ_i^{\text{b}})- P\leq C$ for a
constant $C$ and the corresponding
$\text{SINR}^{\text{b}}_{i,j}\ge \gamma_{i,j}$. An achievable BC
SINR region denoted by $\mathcal{R}_{\text BC}$, is a set
containing all the BC achievable $\bgamma$.
\end{definition}
\begin{definition}\label{MACachievable}
A SINR vector
$\bgamma=(\gamma_{1,1},\ldots,\gamma_{1,N_t},\ldots,\gamma_{K,N_t})$
is said to be achievable for the dual MAC if and only if there
exists a set of $\bQ_1^m,\ldots,\bQ_K^m$ such that
$\sum_{i=1}^{K}\text{tr}(\bQ_i^{\text{m}})\sigma^2-P\leq C$ for a
constant $C$ and the corresponding
$\text{SINR}^{\text{m}}_{i,j}\ge \gamma_{i,j}$. An achievable MAC
SINR region denoted by $\mathcal{R}_{\text MAC}$, is a set
containing all the MAC achievable $\bgamma$.
\end{definition}

In the following corollary, we will show $\mathcal{R}_{\text
MAC}=\mathcal{R}_{\text BC}$.
\begin{Corollary}\label{corollary:sameregion}
For fixed $q_{t}$ and $q_{u}$, and a constant $C$, the MIMO-BC
under the constraint
$q_{t}\sum_{i=1}^{K}\bh_{o}^{\dag}\bQ_i^{\text{b}}\bh_{o}
+q_{u}\sum_{i=1}^{K}\text{tr}(\bQ_i^{\text{b}})- P\leq C$ and the
dual MIMO-MAC under the constraint
$\sum_{i=1}^{K}\text{tr}(\bQ_i^{\text{m}})\sigma^2-P\leq C$
achieve the same SINR region.
\end{Corollary}
\begin{proof}  For
any $\bgamma\in \mathcal{R}_{\text MAC}$, by Definition
\ref{MACachievable}, there exists a set of $\bQ_1^m,\ldots,\bQ_K^m$
such that $\sum_{i=1}^{K}\text{tr}(\bQ_i^{\text{m}})\sigma^2-P\leq
C$ and the corresponding $\text{SINR}^{\text{m}}_{i,j}\ge
\gamma_{i,j}$. It can be readily concluded from Proposition
\ref{thm:Macduality} that there exists a set of
$\bQ_1^b,\ldots,\bQ_K^b$ such that
$q_{t}\sum_{i=1}^{K}\bh_{o}^{\dag}\bQ_i^{\text{b}}\bh_{o}
+q_{u}\sum_{i=1}^{K}\text{tr}(\bQ_i^{\text{b}})- P\leq C$ and the
corresponding $\text{SINR}^{\text{b}}_{i,j}\ge \gamma_{i,j}$. This
implies $\bgamma \in  \mathcal{R}_{\text BC}$. Since $\bgamma$ is an
arbitrary element in $\mathcal{R}_{\text{MAC}}$, we have
$\mathcal{R}_{\text MAC}\subseteq \mathcal{R}_{\text BC}$. In a
similar manner, we have $\mathcal{R}_{\text BC}\subseteq
\mathcal{R}_{\text MAC}$. The proof follows.
\end{proof}

We are now in the position to prove Proposition
\ref{proposition:dual}.

{\it Proof of Proposition \ref{proposition:dual}:} According to
Corollary \ref{corollary:sameregion}, if $C=0$, then under the
constraint $q_{t}\sum_{i=1}^{K}\bh_{o}^{\dag}\bQ_i^{\text{b}}\bh_{o}
+q_{u}\sum_{i=1}^{K}\text{tr}(\bQ_i^{\text{b}})\leq P$ for the BC
and the constraint
$\sum_{i=1}^{K}\text{tr}(\bQ_i^{\text{m}})\sigma^2\leq P$ for the
dual MAC, the two channels have the same SINR region. Since the
achievable rates of user $i$ in the MIMO-MAC and the MIMO-BC are
$r_i^{m}=\sum_{j=1}^{N_r}\log(1+\text{SINR}^{m}_{i,j})$ and
$r_i^{b}=\sum_{j=1}^{N_r}\log(1+\text{SINR}^{b}_{i,j})$, the rate
regions of the two channels are the same. Therefore, Proposition
\ref{proposition:dual} follows. \hfill$\blacksquare$

Note that due to the additional interference constraint, Problem
\ref{eq:objfuntransform} cannot be solved by using the established
duality result in \cite{Viswanath2003:sumcapacity} and
\cite{Jindal03:sumcapacity}, in which only a single sum power
constraint was considered. Our duality result in Proposition
\ref{proposition:dual} can be thought as an extension of the duality
results in
\cite{Viswanath2003:sumcapacity}\cite{Jindal03:sumcapacity} to a
multiple linear constraint case. Moreover, as will be shown in the
following section, our duality result formulates a MIMO-MAC problem
(Problem \ref{eq:objfunMACdual}), which can be efficiently solved.

\section{Dual MAC Weighted Sum Rate Maximization Problem}\label{section:weightsumrateMAC}

In this section, we propose an efficient algorithm to solve Problem
\ref{eq:objfunMACdual}. With the SIC scheme, the achievable rate of
the $k$th user in the dual MIMO-MAC is given by
\begin{equation}
r_k^m=\log\frac{|\bR_{w}+\sum_{j=1}^{k}\bH_j\bQ_j^{\text{m}}\bH_j^{\dag}|}{|\bR_{w}+\sum_{j=1}^{k-1}\bH_j\bQ_j^{\text{m}}\bH_j^{\dag}|}.\label{eq:rateMAC}
\end{equation}
For the MIMO-MAC, the {\it equally} weighted sum rate maximization
is irrespective of the decoding order. However, in general the
weighted sum rate maximization in the MIMO-MAC is affected by the
decoding order. We thus need to consider the optimal decoding order
of the SIC for the dual MIMO-MAC, and further need to consider the
corresponding optimal encoding order of the DPC for the primal BC.

Let $\pi$ be the optimal decoding order, which is a permutation on
the SU index set $\{1,\cdots,K\}$. It follows from
\cite{Tse98:Polymatriod} that the optimal user decoding order $\pi$
for Problem \ref{eq:objfunMACdual} is the order such that
$w_{\pi(1)}\geq w_{\pi(2)}\geq \cdots \geq w_{\pi(K)}$ is satisfied.
The following lemma presents the optimal decoding order of the SIC
for the data streams within a SU (see Appendix \ref{lemma:proof1}
for the proof).
\begin{Lemma}\label{lemma:order}
The optimal data stream decoding order for a particular SU is
arbitrary.
\end{Lemma}

Due to the duality between the MIMO-BC and the MIMO-MAC, for Problem
\ref{Prob:fixq}, the optimal encoding order for the DPC is the
reverse of $\pi$. Because of the arbitrary encoding order for the
data streams within a SU, if we choose a different encoding order
for the BC, the MAC-to-BC mapping algorithm can give different
results which yield the same objective value. Hence, the matrix
$\bQ_i^{\text{b}}$ achieving the optimal objective value are not
unique. With no loss of generality, we assume $w_{1}\geq w_{2}\geq
\cdots \geq w_{K}$ for notational convenience.

According to \eqref{eq:rateMAC}, the objective function of Problem
\ref{eq:objfunMACdual} can be rewritten as
\begin{equation}
f(\bQ^{\text{m}}_1,\cdots,\bQ^{\text{m}}_K):=\sum_{i=1}^{K}\Delta_i\log|\bR_{w}+\sum_{j=1}^{i}\bH_j\bQ_j^{\text{m}}\bH_j^{\dag}|,
\end{equation}
where $\Delta_i:=w_i-w_{i+1}$, and $w_{K+1}:=0$. Clearly, Problem
\ref{eq:objfunMACdual} is a convex problem, which can be solved
through standard convex optimization software packages directly.
However, the standard convex optimization software does not exploit
the special structure of the problem, and thus is computationally
expensive. An efficient algorithm was developed to solve a weighted
sum rate maximization problem for the SIMO-MAC in
\cite{Caire06:MISOMACsumrate}. However, since this algorithm just
consider the case where each users has a single data stream, it is
not applicable to our problem. In the following, we develop a primal
dual method based algorithm \cite{matrix_opt_book} to solve this
problem.

We next rewrite Problem \ref{eq:objfunMACdual} as
\begin{equation}
\underset{{\{\sbQ_i^{\text{m}}\}_{i=1}^K:~\sbQ^{\text{m}}_i\succeq
0}}{\ma}~f(\bQ^{\text{m}}_1,\cdots,\bQ^{\text{m}}_K)~~\text{subject
to~}\sum_{i=1}^{K}\text{tr}(\bQ_i^{\text{m}})\leq
P.\label{probform:transformfinal}
\end{equation}
Recall that the positive semi-definiteness of $\bQ_i^{\text{m}}$
is equivalent to the positiveness of the eigenvalues of
$\bQ_i^{\text {m}}$, i.e., $q_{i,j}\geq0$. Correspondingly, the
Lagrangian function is
\begin{equation}
L(\bQ^{\text{m}}_1,\cdots,\bQ^{\text{m}}_K,\lambda,\delta_{i,j})=f(\bQ^{\text{m}}_1,\cdots,\bQ^{\text{m}}_K)-\lambda\big
(\sum_{i=1}^{K}\text{tr}(\bQ_i^{\text{m}})-P\big
)+\sum_{i=1}^{K}\sum_{j=1}^{M_i}\delta_{i,j}q_{i,j},\label{eq:lagr26}
\end{equation}
where $\lambda$ and $\delta_{i,j}$ are Lagrangian multipliers.
According to the KKT conditions of \eqref{probform:transformfinal},
we have
\begin{align}
\frac{\partial{f(\bQ^{\text{m}}_1,\cdots,\bQ^{\text{m}}_K)}}
{\partial{\bQ_i^{\text{m}}}}-\lambda \bI_{N_r}+\sum_{j=1}^{M_i}\delta_{i,j}\bE_{i,j}=0,\label{eq:KKTMAC}\\
\lambda\big(\sum_{i=1}^{K}\text{tr}(\bQ_i^{\text{m}})-P\big)=0,\\
\delta_{i,j}q_{i,j}=0,\label{eq:KKT}
\end{align}
where $\bE_{i,j}:=\partial q_{i,j}/\partial \bQ_i^{\text{m}}$.
Notice that it is not necessary to compute the actual value of
$\delta_{i,j}$ and $\bE_{i,j}$, because if $\delta_{i,j}\neq0$, then
$q_{i,j}=0$. Thus, the semi-definite constraint turns into
$q_{i,j}=[q_{i,j}]^+$. Thus, we can assume $\delta_{i,j}=0$.

The dual objective function of \eqref{probform:transformfinal} is
\begin{align}
g(\lambda)=\underset{\{\sbQ^{\text{m}}_i\}_{i=1}^K:~\sbQ^{\text
m}_i\ge
0}{\ma}L(\bQ^{\text{m}}_1,\cdots,\bQ^{\text{m}}_K,\lambda).\label{eq:dualobjfunMAC}
\end{align}
Because the problem \eqref{probform:transformfinal} is convex, it is
equivalent to the following minimization problem
\begin{align}\label{prob:inner}
\underset{\lambda}{\mi}~g(\lambda)~~\text{subject to~}~\lambda\geq
0.
\end{align}
We outline the algorithm to solve the problem \eqref{prob:inner}. We
choose an initial $\lambda$ and compute the value of $g(\lambda)$
\eqref{eq:dualobjfunMAC}, and then update $\lambda$ according to the
descent direction of $g(\lambda)$. The process repeats until the
algorithm converges.

It is easy to observe that all the users share the same $\lambda$,
and thus $\lambda$ can be viewed as a water level in the water
filling principle. Once $\lambda$ is fixed, the unique optimal set
$\{\bQ^{\text{m}}_1,\ldots,\bQ^{\text{m}}_K\}$ can be obtained
through the gradient ascent algorithm. In each iterative step,
$\bQ_i^{\text{m}}$ is updated sequentially according to its gradient
direction of \eqref{eq:lagr26}. Denote by $\bQ_i^{\text{m}}(n)$ the
matrix $\bQ_i^{\text{m}}$ at the $n$th iteration step. The gradient
of each step is determined by

\begin{equation}\label{eq:nabla}
\nabla^{(n)}_{\sbQ^{m}_{i}}L:=
\frac{\partial{f\big(\bQ^{\text{m}}_1(n),\cdots,\bQ^{\text{m}}_{i-1}(n),\bQ^{\text{m}}_i(n-1),\ldots,\bQ^{\text{m}}_K(n-1)\big)}}{\partial{\bQ_i^{\text{m}}(n-1)}}
-\lambda\bI_{N_r}.
\end{equation}

Thus, $\bQ_{i}^{\text{m}}(n)$ can be updated according to
\[
\bQ_i^{\text{m}}(n)=\left
[\bQ_i^{\text{m}}(n-1)+t\nabla^{(n)}_{\sbQ^{m}_{k}}L\right]^+,
\]
where $t$ is the step size, and the notation $[\bA]^+$ is defined as
$[\bA]^+:=\sum_j[\lambda_j]^+\bv_j\bv_j^{\dag}$ with $\lambda_j$ and
$\bv_j$ being the $j$th eigenvalue and the corresponding eigenvector
of $\bA$ respectively. The gradient in \eqref{eq:nabla} can be
readily computed as
\begin{equation}
\frac{\partial{f(\bQ^{\text{m}}_1,\cdots,\bQ^{\text{m}}_K)}}{\partial{\bQ_k^{\text{m}}}}=\sum_{j=k}^{K}\Delta_j\left
(\bH_k\bF_{j}(\bQ^{\text{m}}_1,\cdots,\bQ^{\text{m}}_K)^{-1}\bH_k^{\dag}\right
),
\end{equation}
where
$\bF_{j}(\bQ^{\text{m}}_1,\cdots,\bQ^{\text{m}}_K):=\bR_{w}+\sum_{i=1}^{j}\bH_i^{\dag}\bQ_i^{\text{m}}\bH_i$.
We next need to determine the optimal $\lambda$. Since the
Lagrangian function $g(\lambda)$ is convex over $\lambda$, the
optimal $\lambda$ can be obtained through the one-dimensional
search. However, because $g(\lambda)$ is not necessarily
differentiable, the gradient algorithm cannot be applied.
Alternatively, the subgradient method can be used to find the
optimal solution. In each iterative step, $\lambda$ is updated
according to the subgradient direction.

\begin{Lemma}\label{lemma:subgraDIPA}
The sub-gradient of $g({\lambda})$ is
$P-\sum_{i=1}^{K}\text{tr}({\bQ}_i^{\text{m}})$, where ${\lambda}\ge
0$, and ${\bQ}_i^{\text{m}},~i=1,\ldots,K$, are the corresponding
optimal covariance matrices for a fixed $\lambda$ in
\eqref{eq:dualobjfunMAC}.
\end{Lemma}
\begin{proof}
The proof is provided in Appendix \ref{lemma:proof2}.
\end{proof}

Lemma \ref{lemma:subgraDIPA} indicates that the value of $\lambda$
should increase, if $\sum_{i=1}^{K}\text{tr}(\bQ_i^{\text{m}})>P$,
and vice versa. We are now ready to present our algorithm for
solving Problem \ref{eq:objfunMACdual}.

\noindent {\underline{\it Decoupled Iterative Power Allocation
(DIPA) Algorithm :}}
\begin{enumerate}
    \item Initialize $\lambda_{\min}$ and $\lambda_{\max}$;
    \item repeat
    \begin{enumerate}
        \item $\lambda=(\lambda_{\min}+\lambda_{\max})/2$
        \item repeat, initialize
    $\bQ_1^{\text{m}}(0),\cdots,\bQ_K^{\text{m}}(0)$, $n=1$ \\
            $~~~~$for $i=1,\cdots,K$\\
            $~~~~~~~~$$\bQ_i^{\text{m}}(n)=\left [\bQ_i^{\text{m}}(n-1)+t\nabla^{(n)}_{Q^{m}_{i}}L\right ]^+$,\\
            $~~~~$end for\\
            $~~~~$$n=n+1$,
        \item until $\bQ_k^{\text{m}}$ for $k=1,\cdots,K$ converge,
        i.e.,
        $\|\nabla^{(n)}_{\sbQ^{m}_{i}}L\|^2\le
        \hat{\epsilon}$ for a small preset $\hat{\epsilon}$.
        \item if $\sum_{i=1}^{K}\text{tr}(\bQ_i^{\text{m}})>P$, then
        $\lambda_{\min}=\lambda$, elseif $\sum_{i=1}^{K}\text{tr}(\bQ_i^{\text{m}})<P$, then
        $\lambda_{\max}=\lambda$;
    \end{enumerate}
    \item until $|\lambda_{\min}-\lambda_{\max}|\leq \epsilon$,
\end{enumerate}
where $\epsilon>0$ is a constant. The following proposition shows
the convergence property of the DIPA algorithm.
\begin{Propo}\label{proposition:DIPA}
The DIPA algorithm converges to an optimal set of the MAC transmit
signal covariance matrices.
\end{Propo}
\begin{proof}
The DIPA algorithm consists of the inner and outer loops. The inner
loop is to compute $\bQ_i^{\text{m}}$ for $i=1,\cdots,K$. In each
iterative step of the inner loop, we update $\bQ_i^{\text{m}}$ by
fixing other $\bQ_j^{\text{m}}$ with $j\neq i$, and compute the
corresponding gradient. The inner loop uses the gradient ascent
algorithm, which converges to the optimal value due to its
nondecreasing property and the convexity of the objective function.
The outer loop is to compute the optimal Lagrangian multiplier
$\lambda$ in \eqref{prob:inner}. Due to the convexity of the dual
objective function \cite{Boyd_optimization_book}, there is a unique
$\lambda$ achieving the optimal solution in \eqref{prob:inner}.
Hence, we can use an efficient one dimensional line bisection search
(\cite{Yuwei2004:IWF_MAC},\cite{Yuwei2006:sumcapacitycomputation}).
\end{proof}

\begin{Remark}
In the previous work on the sum rate maximization
\cite{Yuwei2004:IWF_MAC} \cite{Jindal05:sumpowerMAC}
\cite{Yuwei2006:sumcapacitycomputation}, the covariance matrix of
each user is the same as the single user water-filling covariance
matrix in a point-to-point link with multiuser interference being
treated as noise \cite{Telatar95:CapacityofMIMO}. However, for the
weighted sum rate maximization problem, the optimal solution does
not possess a water-filling structure. Thus, our DIPA algorithm does
not obey the water-filling principle. In Section
\ref{section:simulation}, Example 1 compares the water-filling
algorithm with the DIPA algorithm. Notably, the formulation of
Problem \ref{eq:objfunMACdual} is similar to the weighted sum rate
problem for the dual MIMO MAC in \cite{Liujia:conjgrad}. The
algorithm proposed therein to handle the dual MIMO MAC problem is
based on gradient projection method \cite{matrix_opt_book}. The
difference between our DIPA algorithm and the algorithm in
\cite{Liujia:conjgrad} is just like the difference between the
algorithms in \cite{Jindal05:sumpowerMAC} and
\cite{Yuwei2006:sumcapacitycomputation}.
\end{Remark}

The DIPA algorithm is an efficient algorithm to obtain the optimal
transmit covariance matrix of the dual MIMO MAC (Problem
\ref{eq:objfunMACdual}). Moreover, the optimal solution to Problem
\ref{Prob:fixq} can be obtained via the MAC-to-BC covariance matrix
mapping algorithm presented in the next section.

\section{MAC-to-BC Covariance Matrix Mapping}\label{section:Mapping}

A covariance matrix mapping algorithm was developed in
\cite{Jindal03:sumcapacity}. However, this algorithm works for the
sum rate maximization problem under a single sum power constraint,
and is not applicable to a weighted sum rate problem under multiple
constraints. In the following, we develop a covariance matrix
mapping algorithm, which computes the BC covariance matrices
$\bQ_i^{\text{b}}$ via the dual MAC covariance matrices
$\bQ_i^{\text{m}}$ such that two channels yield a same weighted sum
rate.

In the MIMO-MAC, according to \eqref{eq:eigendecV}, the transmit
beamforming vectors $\bv_{i,j}$ can be obtained by the eigenvalue
decomposition. The corresponding receive beamforming vector at the
BS, $\bu_{i,j}$, is obtained by using the minimum mean square error
(MMSE) algorithm:
\begin{equation}
\bu_{i,j}=a\big(\sum_{k=1}^{i-1}\sum_{l=1}^{N_r}q_{k,l}\bH_k^{\dag}\bv_{k,l}\bv_{k,l}^{\dag}\bH_k+\sum_{l=1}^{j-1}q_{i,l}\bH_i^{\dag}\bv_{i,l}\bv_{i,l}^{\dag}\bH_i+\bR_w\big)^{-1}\bH_i^{\dag}\bv_{i,j},
\end{equation}
where $a$ is a normalized factor such that $||\bu_{i,j}||=1$.
Throughout the proof of Proposition \ref{thm:Macduality}, we can see
that when the same optimal solutions are achieved the primal BC and
the dual MAC share the same beamforming vectors $\bu_{i,j}$ and
$\bv_{i,j}$. Hence, the transmit beamforming vectors of the BC are
just the receive beamforming vectors of the dual MAC, and the
receive beamforming vectors of the BC are the transmit beamforming
vectors of the dual MAC. Thus, to obtain the transmit signal
covariance matrix of SU$_i$ for the BC, we only need to compute the
power allocated to each data stream. Due to Corollary
\ref{corollary:sameregion}, the dual MAC and the BC can achieve the
same SINR region, i.e.,
$\text{SINR}_{i,j}^{\text{b}}=\text{SINR}_{i,j}^{\text{m}}$. Thus,
for the BC, the power allocated to the beamforming direction
$\bu_{i.j}$ can be obtained by
\begin{equation}
p_{i,j}=\frac{\text{SINR}_{i,j}^{\text{m}}\left(\sum_{k=i+1}^{K}\sum_{l=1}^{N_r}p_{k,l}|\bu_{k,l}^{\dag}\bH_i^{\dag}\bv_{i,j}|^2+\sum_{l=j+1}^{N_r}p_{i,l}|\bu_{i,l}^{\dag}\bH_i^{\dag}\bv_{i,j}|^2+\sigma^2\right)}{|\bu_{i,j}^H\bH_i^{\dag}\bv_{i,j}|^2}.\label{eq:formularPij}
\end{equation}
For the BC, the encoding order is the reverse of the decoding order
of the MAC. Thus, $p_{K,N_r}$ is computed first, $p_{K,N_{r}-1}$ is
computed second, and so on, in the decreasing order of the data
stream index and the user index.

After computing the power for all the beamforming vectors, we obtain
the signal covariance matrix from the BS to SU$_i$,
$\bQ_i^{\text{b}}=\sum_{j=1}^{N_r}p_{i,j}\bu_{i,j}\bu_{i,j}^{\dag}$.
The aforedescribed process can be summarized as the following
algorithm.

\noindent {\underline{\it MAC-to-BC Covariance Matrix Mapping
Algorithm:}}
\begin{enumerate}
    \item Compute $q_{i,j}$ and $\bv_{i,j}$ through eigenvalue
    decomposition:
    $\bQ_i^{\text{m}}=\bV_i{\bf\Lambda}_i\bV^{\dag}_i=\sum_{j=1}^{N_r}q_{i,j}\bv_{i,j}\bv_{i,j}^{\dag}$;
    \item Use the MMSE algorithm to obtain the optimal receiver
    beamforming vector $\bu_{i,j}$ and $\text{SINR}_{i,j}^m$;
    \item Compute $p_{i,j}$ through
    \eqref{eq:formularPij}
    according to the duality between the BC and the MAC;
    \item Compute $\bQ_i^{\text{b}}=\sum_{j=1}^{N_r}p_{i,j}\bu_{i,j}\bu_{i,j}^{\dag}.$
\end{enumerate}
It should be noted that even though an explicit algorithm is not
given, the paper \cite{Yuwei07:perantconst} has mentioned the idea
behind the above algorithm. The MAC-to-BC covariance matrix mapping
allows us to obtain the optimal BC covariance matrices for Problem
\ref{Prob:fixq} by solving Problem \ref{eq:objfunMACdual}.

\section{A Complete Solution to the CR MIMO-BC Weighted Sum Rate Problem}\label{section:overallalgorithm}

We are now ready to present a complete algorithm to solve Problem
\ref{eq:objfuntransform}. The Lagrangian dual objective function
of Problem \ref{eq:objfuntransform} can be rewritten as follows
\begin{align}
g(q_{t},q_{u})&=\underset{{\{\sbQ_i^{\text{b}}\}_{i=1}^K:\sbQ^{\text{b}}_i\succeq
0}}{\ma}~\sum_{i=1}^{K}w_ir_i^{\text{b}}, \label{eq:Lqthqmax}
\end{align}
where the maximization is subject to the constraint
$q_{t}\big(\sum_{i=1}^{K}\bh_{o}^{\dag}\bQ_i^{\text{b}}\bh_{o}
-P_t\big)+q_{u}\big(\sum_{i=1}^{K}\text{tr}(\bQ_i^{\text{b}})-P_{\text
u}\big)\leq 0$. Problem \ref{eq:objfuntransform} is equivalent to
the following problem
\begin{align*}
\underset{q_{t},q_{u}}{\mi}~g(q_{t},q_{u}),~~\text{subject
to~}~q_{t}\geq 0~\text{and}~q_{u}\geq 0.
\end{align*}

Applying the BC-MAC duality in Section \ref{subsection:MACBCdual}
and the DIPA algorithm in Section \ref{section:weightsumrateMAC},
$g(q_{t},q_{u})$ can be obtained. The remaining task is to determine
the optimal $q_{t}$ and $q_{u}$. Since $g(q_{t},q_{u})$ is not
necessarily differentiable, we search the optimal $q_{t}$ and
$q_{u}$ through the subgradient algorithm; that is, in each
iterative step, we update the vector $[q_{t},q_{u}]$ according to
the subgradient direction $\bs=[s_1, s_2]$ of $g(q_{t},q_{u})$.
\begin{Lemma}\label{lemma:subgraSIPA}
The subgradient of $g({q}_{t},{q}_{u})$ is
$\big[P_t-\sum_{i=1}^{K}\bh_{o}^{\dag}{\bQ}_i^{\text{b}}\bh_{o},P_u-\sum_{i=1}^{K}\text{tr}({\bQ}_i^{\text{b}})\big]$,
where ${q}_{t}\ge0,{q}_{u}\ge0$, and
${\bQ}_i^{\text{b}}~,i=1,\ldots,K$, are the corresponding optimal
covariance matrices for the problem \eqref{eq:Lqthqmax}.
\end{Lemma}
\begin{proof}
The proof is given in Appendix \ref{lemma:proof3}.
\end{proof}

It has been shown in \cite{SBoyd03:Subgrad} that with a constant
step size, the subgradient algorithm converges to a value that is
within a small range of the optimal value, i.e.,
\begin{equation}
\lim_{n\rightarrow \infty}|q_{t}^{(n)}-q_{t}^*|<\epsilon,
~\text{and},~\lim_{n\rightarrow \infty}|q_{u}^{(n)}-q_{u}^*|<\epsilon,
\end{equation}
where $q_{t}^*$ and $q_{u}^*$ denote the optimal values, and
$q_{t}^{(n)}$ and $q_{u}^{(n)}$ denote the values of $q_{t}$ and
$q_{u}$ at the $n$th step of the subgradient algorithm,
respectively. This implies that the subgradient method finds an
$\epsilon$-suboptimal point within a finite number of steps. The
number $\epsilon$ is a decreasing function of the step size.
Moreover, if the diminishing step size rule, e.g., the square
summable but not summable step size, is applied, the algorithm is
guaranteed to converge to the optimal value.

We next describe the algorithm to solve Problem
\ref{eq:objfuntransform} as follows.

 \noindent {\underline{\it Subgradient Iterative Power
Allocation (SIPA) Algorithm :}}
\begin{enumerate}
    \item Initialization: ${q^{(1)}_{t}}$, ${q^{(1)}_{u}}$, $n=1$,
    \item repeat
    \begin{enumerate}[2a)]
        \item Find the optimal solution of the dual MAC Problem \ref{eq:objfunMACdual} through the DIPA
        algorithm;
        \item Find the solution of the BC problem \eqref{eq:Lqthqmax} through the MAC-to-BC mapping
        algorithm;
        \item Update $q_{t}^{(n)}$ and $q_{u}^{(n)}$ through a subgradient algorithm
        $q_{t}^{(n+1)}=q_{t}^{(n)}+t(\sum_{i=1}^{K}\bh_o^{\dag}\bQ_i^{\text{b}}\bh_o-P_t)$,
        $q_{u}^{(n+1)}=q_{u}^{(n)}+t(\sum_{i=1}^{K}\text{tr}(\bQ_i^{\text{b}})-P_u)$,
        \item $n=n+1$
    \end{enumerate}
    \item Stop when $|q_{t}^{(n)}(\sum_{i=1}^{K}\bh_o^{\dag}\bQ_i^{\text{b}}\bh_o-P_t)|\leq \epsilon$ and $|q_{u}^{(n)}(\sum_{i=1}^{K}\text{tr}(\bQ_i^{\text{b}})-P_u)|\leq \epsilon$ are satisfied
    simultaneously,
\end{enumerate}
where $t$ denotes the step size of the subgradient algorithm. As a
summary, the flow chart of the SIPA algorithm is depicted in Fig.
\ref{fig:SIPAchart}. We shows that the SIPA algorithm converges to
the optimal solution of Problem \ref{eq:objfun} in the following
proposition.

\begin{Propo}\label{thm:sipaconverge}
The SIPA algorithm converges to the globally optimal solution of
Problem \ref{eq:objfun}.
\end{Propo}
\begin{proof}
The Lagrangian function of Problem \ref{eq:objfun} is given by
\begin{equation}
L(\bQ_1^{\text{b}},\ldots,\bQ_K^{\text{b}},\lambda_1,\lambda_2)
=\sum_{i=1}^{K}w_ir_i^{\text{b}}-\lambda_1\big(\sum_{i=1}^{K}\bh_{o}^{\dag}\bQ_i^{\text{b}}\bh_o-P_t\big)-\lambda_2
\big(\sum_{i=1}^{K}\text{tr}(\bQ_i^{\text{b}})-P_{\text
u}\big)\label{lag:L},
\end{equation}
and the Lagrangian function of Problem \ref{eq:objfuntransform} is
given by
\begin{equation}
L_1(\bQ_1^{\text{b}},\ldots,\bQ_K^{\text{b}},\lambda,q_{t},q_{u})
=\sum_{i=1}^{K}w_ir_i^{\text{b}}-\lambda\big(q_{t}\big(\sum_{i=1}^{K}\bh_{o}^{\dag}\bQ_i^{\text{b}}\bh_o-P_t\big)-q_{u}
\big(\sum_{i=1}^{K}\text{tr}(\bQ_i^{\text{b}})-P_{\text
u}\big)\big).\label{lag:L1}
\end{equation}

Let $\bar{q}_{t}$, $\bar{q}_{u}$, $\bar{\lambda}$, and $\bar{\bQ}_i$
be the optimal values of
$L_1(\bQ_1^{\text{b}},\ldots,\bQ_K^{\text{b}},\lambda,q_{t},q_{u})$,
when the algorithm converges. We thus have
\[
\frac{\partial
L_1(\bQ_1^b,\ldots,\bQ_K^b,\lambda,q_{t},q_{u})}{\partial
\bQ_i^{\text{b}}}\Big|_{\{\bar{\sbQ}_i^b\}_{i=1}^K,\bar{\lambda},\bar{q}_t,\bar{q}_u}=0,
\]
$|\bar{q}_{t}(\sum_{i=1}^{K}\bh_o^{\dag}\bar{\bQ}_i\bh_o-P_t)|= 0$,
and $|\bar{q}_{u}(\sum_{i=1}^{K}\text{tr}(\bar{\bQ}_i)-P_{\text u})|
= 0.$ This means that $\bar{\bQ}_i$ is a locally optimal solution.

According to \eqref{lag:L}, if we select
$\tilde{\lambda}_1=\bar{\lambda}\bar{q}_{t}$,
$\tilde{\lambda}_2=\bar{\lambda}\bar{q}_{u}$, and
$\tilde{\bQ}_i=\bar{\bQ}_i$, then $\tilde{\lambda}_1$,
$\tilde{\lambda}_2$, and $\tilde{\bQ}_i$ satisfy the KKT conditions
of Problem \ref{eq:objfun} and thus are the locally optimal
variables.

Suppose that there exists an optimal set of $\hat{\lambda}_1$,
$\hat{\lambda}_2$, and $\hat{\bQ}_i$ such that
$L(\hat{\bQ}_1,\ldots,\hat{\bQ}_K,\hat{\lambda}_1,\hat{\lambda}_2)>L(\tilde{\bQ}_1,\ldots,\tilde{\bQ}_K,$
$\tilde{\lambda}_1,\tilde{\lambda}_2)$. Clearly, this optimal set of
$\hat{\lambda}_1$, $\hat{\lambda}_2$, and $\hat{\bQ}_i$ satisfy the
KKT conditions of Problem \ref{eq:objfun}. In the sequel, we will
derive a contradiction.

First, we can write
\begin{equation}
L(\tilde{\bQ}_1,\cdots,\tilde{\bQ}_K,\tilde{\lambda}_1,\tilde{\lambda}_2)\geq
L(\hat{\bQ}_1,\cdots,\hat{\bQ}_K,\tilde{\lambda}_1,\tilde{\lambda}_2).\label{ineq:QQ'}
\end{equation}

Suppose that \eqref{ineq:QQ'} does not hold, i.e.,
$L(\tilde{\bQ}_1,\cdots,\tilde{\bQ}_K,\tilde{\lambda}_1,\tilde{\lambda}_2)<
L(\hat{\bQ}_1,\cdots,$
$\hat{\bQ}_K,\tilde{\lambda}_1,\tilde{\lambda}_2)$. Then, according
to the BC-MAC duality in Section \ref{subsection:MACBCdual}, an
objective value of \eqref{eq:macobj} which is larger than
$L(\tilde{\bQ}_1,\cdots,\tilde{\bQ}_K,\tilde{\lambda}_1,\tilde{\lambda}_2)$,
can be found for the fixed $\bar{q}_{t}$ and $\bar{q}_{u}$. However,
from Proposition \ref{proposition:DIPA}, the DIPA algorithm
converges the optimal solution. It is a contradiction.

Secondly, according to the KKT conditions of Problem
\ref{eq:objfun}, we have
\begin{align}
\hat{\lambda}_1\big(\sum_{i=1}^{K}\bh_{o}^{\dag}\hat{\bQ}_i^{\text{b}}\bh_o-P_t\big)=0,\\
\hat{\lambda}_2
\big(\sum_{i=1}^{K}\text{tr}(\hat{\bQ}_i^{\text{b}})-P_{\text
u}\big)=0.
\end{align}
We thus can write:
\begin{equation}
L(\hat{\bQ}_1,\cdots,\hat{\bQ}_K,\tilde{\lambda}_1,\tilde{\lambda}_2)\geq
L(\hat{\bQ}_1,\cdots,\hat{\bQ}_K,\hat{\lambda}_1,\hat{\lambda}_2).\label{ineq:lambda}
\end{equation}
Combining \eqref{ineq:lambda} and \eqref{ineq:QQ'}, we have
\begin{equation}
L(\tilde{\bQ}_1,\cdots,\tilde{\bQ}_K,\tilde{\lambda}_1,\tilde{\lambda}_2)\geq
L(\hat{\bQ}_1,\cdots,\hat{\bQ}_K,\hat{\lambda}_1,\hat{\lambda}_2).
\end{equation}
This contradicts with our previous assumption.
\end{proof}

\begin{Remark}\label{remark:multiPUs}
The algorithm can be extended to the multiple PU case in the
following manner. Assume that there are $N$ PUs. Problem
\ref{eq:objfuntransform} becomes
\begin{align}\label{prob:multiPU}
\begin{split}
\underset{q_{t,j}\geq0,q_{u}\geq 0}{\mi}&\underset{{\{\sbQ_i^{\text{b}}\}_{i=1}^K:~\sbQ^{\text{b}}_i\succeq 0}}{\ma}~\sum_{i=1}^{K}w_ir_i^{\text{b}},\\
\text{subject to~}~\sum_{j=1}^{N}q_{t,j}  \big(
&\sum_{i=1}^{K}\bh_{o,j}^{\dag}\bQ_i^{\text{b}}\bh_{o,j} -P_{t,j}
\big) +q_{u} \big( \sum_{i=1}^K\text{tr}(\bQ_i^{\text{b}})-P_u \big)
\leq 0,
\end{split}
\end{align}
where $q_{t,j}$ is the auxiliary variable for the $j$th PU,
$\bh_{o,j}$ is the channel response from the BS to the $j$th PU, and
$P_{t,j}$ is the interference threshold of the $j$th PU. The role of
auxiliary variables $q_{t,j}$ is similar to that of $q_t$ in the
single PU case. It is thus straightforward to modify the SIPA
algorithm to solve the problem for the multiple PU case.  Moreover,
it should be noted that the multiple interference constraints of the
problem \eqref{prob:multiPU} can be transformed to the per-antenna
power constraints \cite{Yuwei07:perantconst} by setting $\bh_{o,j}$,
$j=1,\cdots,N_t$, to be the $j$th column of the identity matrix. Not
limited by the sum rate maximization problem with interference power
constraints, the method proposed in this paper can be easily applied
to solve the transmitter optimization problem (e.g. beamforming
optimization) for MIMO BC with multiple arbitrary linear power
constraints.
\end{Remark}

\section{Simulation Results}\label{section:simulation}
In this section, we provide the simulation results to show the
effectiveness of the proposed algorithm. In the simulations, for
simplicity, we assume that the BS is at the same distance, $l_1$, to
all SUs, and the same distance, $l_2^{(n)}$, to PU$_n$. In the
single PU case, we will drop the superscript and simply use notation
$l_2$. Suppose that the same path loss model can be used to describe
the transmissions from the BS to the SUs and to the PUs, and the
pass loss exponent is 4. The elements of matrix $\bH$ are assumed to
be circularly symmetric complex Gaussian (CSCG) RVs with mean zero
and variance 1, and $\bh_o$ can be modeled as
$\bh_o=(l_1/l_2)^2\ba_n$, where $\ba_n$ is a $N_t\times 1 $ vector
whose elements are CSCG RVs with mean zero and variance 1. The noise
covariance matrix at the BS is assumed to be the identity matrix,
and the sum power and interference power are defined in dB relative
to the noise power, and $P_t$ is chosen to be $0$ dB. For all cases,
we choose $l_1=l_2$, except for explicitly stated.

\begin{example} In Fig. \ref{fig:compwf}, we examine the validity of
the DIPA algorithm. In this example, we choose $K=1$ (a single SU
case), $N_t=4$, $N_r=4$, and $P_u=10$ dB. It is well known that the
optimal transmit signal covariance matrix can be obtained through
the water-filling principle \cite{Telatar95:CapacityofMIMO}. As can
be observed from Fig. \ref{fig:compwf}, in several iterations, the
DIPA algorithm converges to the optimal solution obtained by using
the water-filling principle.
\end{example}

\begin{example}
In Fig. \ref{fig:dipaconv}, we show the convergence property of the
DIPA algorithm. In this example, we choose $K=20$ and $P_u=10$ dB.
It can be observed from this figure that the algorithm converges to
the optimal solution within several iteration steps.
\end{example}

\begin{example} In Figs. \ref{fig:rate} and \ref{fig:power}, we consider
a SU MIMO-BC network with $K=5$, $N_t=5$, $N_r=3$, and $P_u=13$ dB.
In this example, the SUs with $w_1=5$ and $w_i=1,~i=2,\ldots,K$ are
assumed to share the same spectrum band with two PUs. Fig.
\ref{fig:rate} plots the weighted sum rate versus the number of
iterations of the SIPA algorithm for step sizes $t=0.1$ and
$t=0.01$. As can be seen from the figure, the step size affects the
accuracy and convergence speed of the algorithm. Fig.
\ref{fig:power} plots the sum power at the BS and the interference
power at the PUs versus the number of iterations. It can be seen
from the figure that the sum power and the interference power
approach to $P_u=13$ dB and $P_t=0$ dB respectively when the SIPA
algorithm converges. This implies that the sum power and
interference constraints are satisfied with equalities when the SIPA
algorithm converges.
\end{example}

\begin{example} Fig. \ref{fig:sumpowerincr} plots the achievable sum
rates versus the sum power in the single PU case and the case with
no PU. We choose $K=5$, $N_t=5$, and $N_r=3$.  As can be seen from
Fig. \ref{fig:sumpowerincr}, in the low sum power regime, the
achievable sum rate in the case with no PU is quite close to the one
in the single PU case while in the high sum power regime, the
achievable sum rate in the case with no PU is much higher than the
one in the single PU case. This is because the additional constraint
reduces the degrees of freedom of the system.

\end{example}

\begin{example} In this example, we consider the influence of the interference constraint
on the achievable sum rate of the SUs. In this example, $N_t=5$,
$K=5$, and $N_r=3$. The sum power constraint for the BS is assumed
to be 15 dB and 20 dB. Fig. \ref{fig:dist} compares the sum rate
achieved in a PU case with one achieved in the case with no PU as
$l_2/l_1$ varies from 1 to 12. It can be observed from the figure
that the achievable sum rate increases as the PU moves away from the
BS, and the influence of the PU reduces to zero after the $l_2/l_1$
is larger than a certain threshold.
%
%interference constraint will not influence the achievable sum rate
%when PU is far away for the BS.
\end{example}

\section{Conclusions}\label{section:conclusion}
In this paper, we developed a new BC-MAC duality result, which can
be viewed as an extension of existing dual results developed under
either a sum power constraint or per-antenna power constraints.
Exploiting this duality result, we proposed an efficient algorithm
to solve the CR MIMO-BC weighted sum rate maximization problem. We
further showed that the proposed algorithm converges to the globally
optimal solution.

\def\appref#1{Appendix~\ref{#1}}
\appendix
\renewcommand{\thesubsection}{\Alph{subsection}}
\makeatletter
\renewcommand{\subsection}{%
\@startsection {subsection}{2}{\z@ }{2.0ex plus .5ex minus .2ex}%
{-1.0ex plus .2ex}{\it }} \makeatother

\subsection{Lemma \ref{lemma:optimalcondition} and its proof}

The following lemma describes an important property that will be
used in the proof of other lemmas.
\begin{Lemma}\label{lemma:optimalcondition}
For fixed $q_{t}$ and $q_{u}$, the maximum weighted sum rate in
\eqref{eq:prob3obj} is achieved when the constraint
\eqref{eq:transformBCcons} is satisfied with equality.
\end{Lemma}

\begin{proof}
We here adopt the DPC scheme, which is a capacity achieving strategy
for the MIMO-BC \cite{Shamai06:MIMOBCcapacity}. Let the permutation
$\pi$ represent the encoding order when the optimal solution is
achieved. Assume that SU$_{\pi(1)}$ is encoded first such that the
signal of SU$_{\pi(1)}$ is noncausally known to the BS before the
signals from the other SUs are encoded. Thus, in the DPC scheme the
signal from SU$_{\pi(1)}$ has no impact on the rates achieved by the
other SUs. We prove this lemma by contradiction.

Suppose that $\bQ^{\text{b}}_{\pi(1)}$ is the optimal signal
covariance matrix of SU$_{\pi(1)}$. Assume that the constraint
\eqref{eq:transformBCcons} is satisfied with a strict inequality
when the optimal solution is achieved. Thus, we can always find an
$\epsilon>0$ such that
\begin{align}
&q_{t}\big
(\sum_{i=2}^{K}\bh_{o}^{\dag}(\bQ^{\text{b}}_{\pi(i)})\bh_{o}+\bh_{o}^{\dag}(\bQ^{\text{b}}_{\pi(1)}+\epsilon
\bI)\bh_{o} -P_t\big )+q_{u}\big
(\sum_{i=2}^{K}\text{tr}(\bQ^{\text{b}}_{\pi(i)})+\text{tr}(\bQ^{\text{b}}_{\pi(1)}+\epsilon
\bI)-P_u\big
)\notag\\
&=q_{t}\big
(\sum_{i=1}^{K}\bh_{o}^{\dag}(\bQ^{\text{b}}_{\pi(i)})\bh_{o}+\bh_{o}^{\dag}(\epsilon
\bI)\bh_{o} -P_t\big )+q_{u}\big
(\sum_{i=1}^{K}\text{tr}(\bQ^{\text{b}}_{\pi(i)})+\text{tr}(\epsilon
\bI)-P_u\big )< 0.\label{eq:lemproof1}
\end{align}
Moreover, the rate achieved by user $\pi(1)$ in the MIMO-BC can be
written as
\begin{equation*}
r_{\pi(1)}^{\text{b}}=\log\frac{\Big|\bI+\sum_{i=1}^{K}\bH_{\pi(1)}\bQ_{\pi(i)}^{\text{b}}\bH_{\pi(1)}^{\dag}\Big|}{\Big|\bI+\sum_{i=2}^{K}\bH_{\pi(1)}\bQ_{\pi(i)}^{\text{b}}\bH_{\pi(1)}^{\dag}\Big|}.
\end{equation*}
Due to the positive semi-definiteness property of
$\bQ_i^{\text{b}}$, we have
\begin{align}
&\log\Big|\bI+\sum_{i=2}^{K}\bH_{\pi(1)}\bQ^{\text{b}}_{\pi(i)}\bH_{\pi(1)}^{\dag}+\bH_{\pi(1)}(\bQ^{\text{b}}_{\pi(1)}+\epsilon
\bI)\bH_{\pi(1)}^{\dag}\Big|\notag\\
=&\log\Big|\bI+\sum_{i=2}^{K}\bH_{\pi(1)}\bQ^{\text{b}}_{\pi(i)}\bH_{\pi(1)}^{\dag}\Big|\notag\\
+&\log\Big|\bI\!\!+(\bI\!+\sum_{i=2}^{K}\bH_{\pi(1)}\bQ^{\text{b}}_{\pi(i)}\bH_{\pi(1)}^{\dag})^{-1/2}
\bH_{\pi(1)}(\bQ^{\text{b}}_{\pi(1)}\!\!+\epsilon
\bI)\bH_{\pi(1)}^{\dag}(\bI\!+\sum_{i=2}^{K}\bH_{\pi(1)}\bQ^{\text{b}}_{\pi(i)}\bH_{\pi(1)}^{\dag})^{-1/2}\Big|\notag
\end{align}
\begin{align}
=&\log\Big|\bI+\sum_{i=2}^{K}\bH_{\pi(1)}\bQ^{\text{b}}_{\pi(i)}\bH_{\pi(1)}^{\dag}\Big|+\log\Big|\bI+\bG^{\dag}(\bQ^{\text{b}}_{\pi(1)}+\epsilon
\bI)\bG\Big|\notag \\
=&\log\Big|\bI+\sum_{i=2}^{K}\bH_{\pi(1)}\bQ^{\text{b}}_{\pi(i)}\bH_{\pi(1)}^{\dag}\Big|+\log\Big|\bI+{\bf \Lambda}+\epsilon {\bf \Sigma}\Big|\label{eq:diag}\\
>&\log\Big|\bI+\sum_{i=2}^{K}\bH_{\pi(1)}\bQ^{\text{b}}_{\pi(i)}\bH_{\pi(1)}^{\dag}\Big|+\log\Big|\bI+{\bf
\Lambda}\Big|\notag \\
=&\log\Big|\bI+\sum_{i=1}^{K}\bH_{\pi(1)}\bQ^{\text{b}}_{\pi(i)}\bH_{\pi(1)}^{\dag}\Big|,\label{eq:lemproof}
\end{align}
where
$\bG=\bH_{\pi(1)}^{\dag}(\bI+\sum_{i=2}^{K}\bH_{\pi(1)}\bQ^{\text{b}}_{\pi(i)}\bH_{\pi(1)}^{\dag})^{-1/2}$,
and ${\bf \Lambda}$ and ${\bf \Sigma}$ are diagonal matrices. Eq.
\eqref{eq:diag} is due to the fact that the optimal covariance
matrix for a MIMO has the water-filling structure
\cite{Yuwei2004:IWF_MAC}\cite{Telatar95:CapacityofMIMO}, i.e., if we
apply singular value decomposition to $\bG$, $\bG=\bV\bS\bU$, where
$\bV$ and $\bU$ are unitary matrices, and $\bS$ is a diagonal
matrix, then the optimal $\bQ^{\text{b}}_{\pi(1)}$ can be written as
$\bQ^{\text{b}}_{\pi(1)}=\bU^{\dag}\bR\bU$, where $\bR$ is a
diagonal matrix. Thus, we have $\bf\Lambda=\bS\bR\bS$ and
$\bf\Sigma=\bS\bS$.

According to \eqref{eq:lemproof} and \eqref{eq:lemproof1},
$\bQ^{\text{b}}_{\pi(1)}+\epsilon \bI$ is a better solution for the
sum rate problem than $\bQ^{\text{b}}_{\pi(1)}$, which contradicts
with the assumption. Therefore, the constraint must be satisfied
with equality.
\end{proof}

\subsection {Proof of Proposition
\ref{lemma:BC2BC}}\label{proposition:proof1}

The proof consists of two parts. In the first part, we show that
either optimal solution is feasible for both problems. In the second
part, we show that Problem \ref{eq:objfun} and Problem
\ref{eq:objfuntransform} have the same solution.

The Lagrangian function of Problem \ref{eq:objfun} is
\begin{equation}
L_1(\bQ^{\text{b}}_1,\cdots,\bQ^{\text{b}}_K,\lambda_{t},\lambda_{u})=\sum_{i=1}^Kw_ir_i^{\text{b}}-\lambda_{t}\big(\sum_{i=1}^{K}\bh_{o}^{\dag}\bQ_i^{\text{b}}\bh_{o}
-
P_t\big)-\lambda_{u}\big(\sum_{i=1}^{K}\text{tr}(\bQ_i^{\text{b}})-P_u\big),\label{eq:lagranL1}
\end{equation}
where $\lambda_{t}$ and $\lambda_{u}$ are the Lagrangian
multipliers. The optimal objective value is
\begin{equation}\label{eq:49}
\underset{\lambda_{t},\lambda_{u}}{\mi}\underset{\sbQ^{\text{b}}_1,\cdots,\sbQ^{\text{b}}_K}{\ma}L_1(\bQ^{\text{b}}_1,\cdots,\bQ^{\text{b}}_K,\lambda_{t},\lambda_{u}).
\end{equation}
Assume the optimal variables are $\bar{\lambda}_{t}$,
$\bar{\lambda}_{u}$ and
$\bar{\bQ}^{\text{b}}_1,\cdots,\bar{\bQ}^{\text{b}}_K$, and the
corresponding optimal value is $\bar{C}$.

The Lagrangian function of Problem \ref{eq:objfuntransform} is:
\begin{equation}
L_2(\bQ^{\text{b}}_1,\cdots,\bQ^{\text{b}}_K,q_{t},q_{u},\lambda)=\sum_iw_ir_i^{\text{b}}-\lambda\Big(q_{t}\big(\sum_{i=1}^{K}\bh_{o}^{\dag}\bQ_i^{\text{b}}\bh_{o}
- P_t\big)+q_{u}\big(\sum_{i=1}^{K}\text{tr}(\bQ_i^{\text{b}})-
P_u\big)\Big),\label{eq:lagranL2}
\end{equation}
where $\lambda$ is the Lagrangian multiplier. The optimal objective
value is
\begin{equation}\label{eq:51}
\underset{q_{t},q_{u},\lambda}{\mi}\underset{\sbQ^{\text{b}}_1,\cdots,\sbQ^{\text{b}}_K}{\ma}L_2(\bQ^{\text{b}}_1,\cdots,\bQ^{\text{b}}_K,q_{t},q_{u},\lambda).
\end{equation}

Suppose that the optimal variables are $\tilde{q}_{t}$,
$\tilde{q}_{u}$, $\tilde{\lambda}$, and
$\tilde{\bQ}^{\text{b}}_i~,i=1,\ldots,K$, and the corresponding
optimal objective value is $\tilde{C}$. We just need to prove
$\bar{C}=\tilde{C}$.

We now present the first part of the proof. According to the KKT
condition of Problem \ref{eq:objfuntransform}, we have
\begin{align}
\frac{\partial
L_2(\tilde{\lambda},\tilde{\bQ}^{\text{b}}_1,\cdots,\tilde{\bQ}^{\text{b}}_K,\tilde{q}_{t},\tilde{q}_{u})}{\partial
q_{t}}=\tilde{\lambda}\big(\sum_{i=1}^{K}\bh_{o}^{\dag}\tilde{\bQ}_i^{\text{b}}\bh_{o}
- P_t\big)=0,\label{eq:lem2prof} \\
\frac{\partial
L_2(\tilde{\lambda},\tilde{\bQ}^{\text{b}}_1,\cdots,\tilde{\bQ}^{\text{b}}_K,\tilde{q}_{t},\tilde{q}_{u})}{\partial
q_{u}}=\tilde{\lambda}\big(\sum_{i=1}^{K}\text{tr}(\tilde{\bQ}_i^{\text{b}})-
P_u\big)=0.\label{eq:lem2prof1}
\end{align}
Recall that the Lagrangian multiplier $\tilde{\lambda}$ is
non-negative. Furthermore, if $\tilde{\lambda}=0$, we have
$\tilde{q}_{t}(\sum_{i=1}^{K}\bh_{o}^{\dag}\tilde{\bQ}_i\bh_{o}
-P_t)+\tilde{q}_{u}(\sum_{i=1}^{K}\text{tr}(\tilde{\bQ}_i)-P_u)< 0$
from the KKT conditions. This contradicts with Lemma
\ref{lemma:optimalcondition}. Thus, we always have
$\tilde{\lambda}>0$ and can readily conclude that
$\sum_{i=1}^{K}\bh_{o}^{\dag}\tilde{\bQ}_i^{\text{b}}\bh_{o} = P_t$
and $\sum_{i=1}^{K}\text{tr}(\tilde{\bQ}_i^{\text{b}}) = P_u$ are
satisfied simultaneously. The optimal solution of Problem
\ref{eq:objfuntransform} is also a feasible solution of Problem
\ref{eq:objfun}. On the other hand, it is obvious that the feasible
solution for Problem \ref{eq:objfun} is also the feasible solution
for Problem \ref{eq:objfuntransform}.

We next prove the second part by using contradiction. Let us first
suppose $\bar{C}>\tilde{C}$. For \eqref{eq:lagranL2}, if we select
$\bQ^{\text{b}}_i=\bar{\bQ}^{\text{b}}_i$ for $i=1,\ldots,K$,
$\lambda=1$, $q_{t}=\bar{\lambda}_{t}$ and
$q_{u}=\bar{\lambda}_{u}$, then $L_2=\bar{C}>\tilde{C}$. It
contradicts to the fact that $\tilde{C}$ is the optimal objective
value for \eqref{eq:51}.

We now assume $\bar{C}<\tilde{C}$. Recall that $\tilde{\lambda}\neq
0$, for \eqref{eq:lagranL2}. If we select
$\bQ^{\text{b}}_i=\tilde{\bQ}^{\text{b}}_i$ for $i=1,\ldots,K$,
$\lambda_{t}=\tilde{\lambda}\tilde{q_{t}}$ and
$\lambda_{u}=\tilde{\lambda}\tilde{q}_{u}$, then
$L_1=\tilde{C}>\bar{C}$, which contradicts with the fact that
$\bar{C}$ is the optimal objective value for \eqref{eq:49}.

Therefore, the optimal solutions for Problem
\ref{eq:objfuntransform} and Problem \ref{eq:objfun} are the same.
\hfill $\blacksquare$

\subsection{Proof of Lemma \ref{lemma:order}}\label{lemma:proof1}
According to previous discussions, the signal from each SU is
divided into several data streams. We now show that the optimal
encoding order of these data streams are arbitrary. It is well known
that the optimal objective value of the MAC equally weighted sum
rate problem can be achieved by adopting any ordering
\cite{Yuwei2004:IWF_MAC}\cite{Jindal05:sumpowerMAC}\cite{Yuwei2006:sumcapacitycomputation};
that is, when all the users have the same weights, the optimal
solution of the weighted sum rate maximization problem is
independent of the decoding order. Analogously, the data streams
within a SU share the same weight. Thus, an arbitrary encoding order
of those data streams within a SU can achieve the optimal solution.
\hfill $\blacksquare$

\subsection{Proof of Lemma
\ref{lemma:subgraDIPA}}\label{lemma:proof2} Let $s$ be the
sub-gradient of $g(\tilde{\lambda})$.  For a given
$\tilde{\lambda}\ge0$, the subgradient $s$ of $g(\tilde{\lambda})$
satisfies $g(\check{\lambda})\geq
g(\tilde{\lambda})+s(\check{\lambda}-\tilde{\lambda})$, where
$\check\lambda$ is any feasible value. Let
$\check{\bQ}^{\text{m}}_i,~i=1,\ldots,K$, be the optimal covariance
matrices in \eqref{eq:dualobjfunMAC} for $\lambda=\check{\lambda}$,
and $\tilde{\bQ}^{\text{m}}_i,~i=1,\ldots,K$, be the optimal
covariance matrices in \eqref{eq:dualobjfunMAC} for
$\lambda=\tilde{\lambda}$. We express $g(\check{\lambda})$ as
%\begin{eqnarray*}
\begin{align*}
g(\check{\lambda})&=\underset{\sbQ^{\text{m}}_1,\cdots,\sbQ^{\text{m}}_K}{\ma}\Big
(f(\bQ^{\text{m}}_1,\cdots,\bQ^{\text{m}}_K)-\check{\lambda}(\sum_{i=1}^{K}\text{tr}(\bQ_i^{\text{m}})-P)\Big)\\
&=f(\check{\bQ}^{\text{m}}_1,\cdots,\check{\bQ}^{\text{m}}_K)-\check{\lambda}\Big(\sum_{i=1}^{K}\text{tr}(\check{\bQ}^{\text{m}}_i)-P\Big)\\
&\geq
f(\tilde{\bQ}^{\text{m}}_1,\cdots,\tilde{\bQ}^{\text{m}}_K)-\check{\lambda}\Big(\sum_{i=1}^{K}\text{tr}(\tilde{\bQ}_i^{\text{m}})-P\Big)\\
%\end{eqnarray*}
%\begin{align*}
&=f(\tilde{\bQ}^{\text{m}}_1,\cdots,\tilde{\bQ}^{\text{m}}_K)\!-\!\tilde{\lambda}\Big(\sum_{i=1}^{K}\text{tr}(\tilde{\bQ}_i^{\text{m}})\!\!-\!P\Big)\!\!+\!
\tilde{\lambda}\Big(\sum_{i=1}^{K}\text{tr}(\tilde{\bQ}_i^{\text{m}})\!-\!P\Big)\!-\!\check{\lambda}\Big(\sum_{i=1}^{K}\text{tr}(\tilde{\bQ}_i^{\text{m}})-P\Big)\notag
\\
&=g(\tilde{\lambda})+\Big(P-\sum_{i=1}^{K}\text{tr}(\tilde{\bQ}_i^{\text{m}})\Big)(\check{\lambda}-\tilde{\lambda}),
\end{align*}
where $s:=P-\sum_{i=1}^{K}\text{tr}(\tilde{\bQ}_i^{\text{m}})$ is
the subgradient of $g(\tilde{\lambda})$. This concludes the proof.
\hfill $\blacksquare$

\subsection{Proof of Lemma
\ref{lemma:subgraSIPA}}\label{lemma:proof3}
The subgradient $\bs$ of $g(\tilde{q}_{t},\tilde{q}_{u})$ satisfies
%\begin{equation*}
$g(\bar{q}_{t},\bar{q}_{u})\geq
g(\tilde{q}_{t},\tilde{q}_{u})+([\bar{q}_{t},\bar{q}_{u}]-[\tilde{q}_{t},\tilde{q}_{u}])\cdot
\bs^T$,
%\end{equation*}
where $[\bar{q}_{t},\bar{q}_{u}]$ is any feasible vector. Let
$\bar{\bQ}_i^{\text{b}}~i=1,\ldots,K,$ be the optimal matrices of
the problem \eqref{eq:Lqthqmax} for $q_{t}=\bar{q}_{t}$ and
$q_{u}=\bar{q}_{u}$, and let
$\tilde{\bQ}_i^{\text{b}}~i=1,\ldots,K,$ be the optimal matrices of
the problem \eqref{eq:Lqthqmax} for $q_{t}=\tilde{q}_{t}$ and
$q_{u}=\tilde{q}_{u}$. We express $g(\bar{q}_{t},\bar{q}_{u})$ as
\begin{align}
&g(\bar{q}_{t},\bar{q}_{u})\!=\underset{\sbQ^{\text{b}}_1,\cdots,\sbQ^{\text{b}}_K}{\ma}\!\sum_{i=1}^{M}w_ir_i^{\text{b}}\\
=&\!\sum_{i=1}^{M}w_i\bar{r}_i^{\text{b}}\!-\!\bar{\lambda}\big(\bar{q}_{t}\big(\sum_{i=1}^{K}\bh_{o}^{\dag}\bar{\bQ^{\text{b}}}_i\bh_{o}
-P_t\big)+\bar{q}_{u}\big(\sum_{i=1}^{K}\text{tr}(\bar{\bQ^{\text{b}}}_i)-P_u\big)\big)\label{eq:lemma5p1}
\\
\geq&
\sum_{i=1}^{M}w_i\tilde{r}_i^{\text{b}}-\bar{\lambda}\Big(\bar{q}_{t}\big(\sum_{i=1}^{K}\bh_{o}^{\dag}\tilde{\bQ}_i^{\text{b}}\bh_{o}
-P_t\big)+\bar{q}_{u}\big(\sum_{i=1}^{K}\text{tr}(\tilde{\bQ}_i^{\text{b}})-P_u\big)\Big)\label{ineq:duetoopt}\\
=&\sum_{i=1}^{M}w_i\tilde{r}_i^{\text{b}}-\tilde{\lambda}\Big(\tilde{q}_{t}\big(\sum_{i=1}^{K}\bh_{o}^{\dag}\tilde{\bQ}_i^{\text{b}}\bh_{o}
-P_t\big)+\tilde{q}_{u}\big(\sum_{i=1}^{K}\text{tr}(\tilde{\bQ}_i^{\text{b}})-P_u\big)\Big)\notag\\
&+\tilde{\lambda}\!\Big(\!\tilde{q}_{t}\big(\!\sum_{i=1}^{K}\bh_{o}^{\dag}\tilde{\bQ}_i^{\text{b}}\bh_{o}
\!\!-\!\!P_t\big)\!\!+\!\!\tilde{q}_{u}\big(\!\sum_{i=1}^{K}\text{tr}(\tilde{\bQ}_i^{\text{b}})\!\!-\!\!P_u\big)\!\Big)\!
-\!\bar{\lambda}\!\Big(\!\bar{q}_{t}\big(\!\sum_{i=1}^{K}\bh_{o}^{\dag}\tilde{\bQ}_i^{\text{b}}\bh_{o}
\!\!-\!\!P_t\big)\!\!+\!\!\bar{q}_{u}\big(\!\sum_{i=1}^{K}\text{tr}(\tilde{\bQ}_i^{\text{b}})\!\!-\!\!P_u\big)\!\Big)\notag
\end{align}
\begin{align}
=&g(\tilde{q}_{t},\tilde{q}_{u})+\Big(\sum_{i=1}^{K}\bh_{o}^{\dag}\tilde{\bQ}_i^{\text{b}}\bh_{o}
-P_t\Big)(\tilde{\lambda} \tilde{q}_{t}-\bar{\lambda} \bar{q}_{t})+\Big(\sum_{i=1}^{K}\text{tr}(\tilde{\bQ}_i^{\text{b}})-P_u\Big)(\tilde{\lambda} \tilde{q}_{u}-\bar{\lambda} \bar{q}_{u})\notag\\
=&g(\tilde{q}_{t},\tilde{q}_{u})+\Big(\sum_{i=1}^{K}\bh_{o}^{\dag}\tilde{\bQ}_i^{\text{b}}\bh_{o}
-P_t\Big)(\tilde{\lambda} \tilde{q}_{t}-\bar{\lambda} \tilde{q}_{t}+\bar{\lambda} \tilde{q}_{t}-\bar{\lambda} \bar{q}_{t})\notag\notag\\
&+\Big(\sum_{i=1}^{K}\text{tr}(\tilde{\bQ}_i^{\text{b}})-P_u\Big)(\tilde{\lambda}
\tilde{q}_{u}-\bar{\lambda} \tilde{q}_{u}+\bar{\lambda}
\tilde{q}_{u}-\bar{\lambda} \bar{q}_{u})\notag \\
=&g(\tilde{q}_{t},\tilde{q}_{u})+\Big(\sum_{i=1}^{K}\bh_{o}^{\dag}\tilde{\bQ}_i^{\text{b}}\bh_{o}
-P_t\Big)(\tilde{\lambda} \tilde{q}_{t}-\bar{\lambda}
\tilde{q}_{t})+\Big(\sum_{i=1}^{K}\bh_{o}^{\dag}\tilde{\bQ}_i^{\text{b}}\bh_{o}
-P_t\Big)(\bar{\lambda} \tilde{q}_{t}-\bar{\lambda}
\bar{q}_{t})\notag \\
&+\Big(\sum_i\text{tr}(\tilde{\bQ}_i^{\text{b}})-P_u\Big)(\tilde{\lambda}
\tilde{q}_{u}-\bar{\lambda}
\tilde{q}_{u})+\Big(\sum_{i=1}^{K}\text{tr}(\tilde{\bQ}_i^{\text{b}})-P_u\Big)(\bar{\lambda}
\tilde{q}_{u}-\bar{\lambda} \bar{q}_{u})\notag \\
=&g(\tilde{q}_{t},\tilde{q}_{u})+\Big(\sum_{i=1}^{K}\bh_{o}^{\dag}\tilde{\bQ}_i^{\text{b}}\bh_{o}
-P_t\Big)(\bar{\lambda} \tilde{q}_{t}-\bar{\lambda}
\bar{q}_{t})+\Big(\sum_{i=1}^{K}\text{tr}(\tilde{\bQ}_i^{\text{b}})-P_u\Big)(\bar{\lambda}
\tilde{q}_{u}-\bar{\lambda}
\bar{q}_{u})\label{eq:duetolemma1}\\
=&g(q_{t},q_{u})+\bar{\lambda}([\bar{q}_{t},\bar{q}_{u}]-[\tilde{q}_{t},
\tilde{q}_{u}])\cdot \bs^T,\notag
\end{align}
where
$\bs:=[P_t-\sum_{i=1}^{K}\bh_{o}^{\dag}\tilde{\bQ}_i^{\text{b}}\bh_{o}
,P_u-\sum_{i=1}^{K}\text{tr}(\tilde{\bQ}_i^{\text{b}})]$. Eq.
\eqref{eq:lemma5p1} is due to the fact that the dual objective
function of the problem \eqref{eq:Lqthqmax}, and
$\bar{r}_i^{\text{b}}$, $\bar{\lambda}$, and
$\bar{\bQ}_i^{\text{b}}$ are the optimal variables for the fixed
$\bar{q}_{t}$ and $\bar{q}_{u}$. The inequality
\eqref{ineq:duetoopt} is because
$\bar{\bQ}_i^{\text{b}}~,i=1,\ldots,K$, are the optimal signal
covariance matrices for the fixed $\bar{q}_{t}$ and $\bar{q}_{u}$.
The equality \eqref{eq:duetolemma1} is due to Lemma
\ref{lemma:optimalcondition}. Thus, $\bs$ is the subgradient of
$g(\tilde{q}_{t},\tilde{q}_{u})$. \hfill $\blacksquare$

{\linespread{1.2}{

\newpage
\begin{figure}
      \centering
      \psfrag{h1}{$\bH_{1}$}
      \psfrag{h2}{$\bH_{2}$}
      \psfrag{hk}{$\bH_{K}$}
      \psfrag{u1}{$SU_{1}$}
      \psfrag{u2}{$SU_{2}$}
      \psfrag{uk}{$SU_{K}$}
      \psfrag{g}{$\bh_{o}$}
      \psfrag{PU}{$\text{PU}$}
      \psfrag{BS}{$\text{BS}$}
     \includegraphics[width = 95mm,height=40mm]{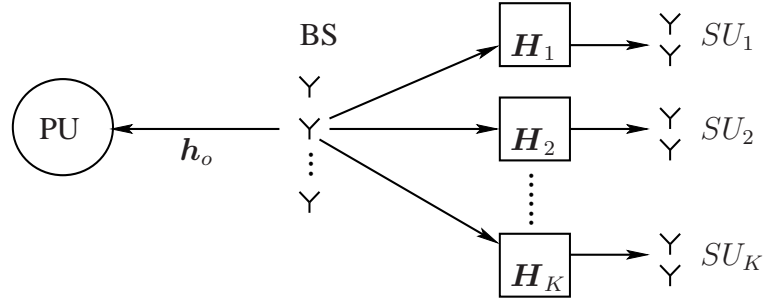}
     \caption{The system model for MIMO-BC based cognitive radio networks.
     There are $K$ SUs. The BS of the SUs has $N_t$ transmit antennas,
     and each SU is equipped with $N_r$ receive antennas.}
     \label{fig:sysmodel}
\end{figure}

\begin{figure}
    \begin{minipage}[b]{.48\linewidth}
        \centering
        \psfrag{h1}{$\bH_{1}$}
        \psfrag{h2}{$\bH_{2}$}
        \psfrag{hk}{$\bH_{K}$}
        \psfrag{y1}{$\by_{1}$}
        \psfrag{y2}{$\by_{2}$}
        \psfrag{yk}{$\by_{K}$}
        \psfrag{y}{$\bx$}
        \psfrag{z1}{$\bz_1$}
        \psfrag{z2}{$\bz_2$}
        \psfrag{zk}{$\bz_k$}
        \psfrag{BS}{$\text{BS}$}
        \vspace{8mm}
        \includegraphics[width = 80mm,height=52mm]{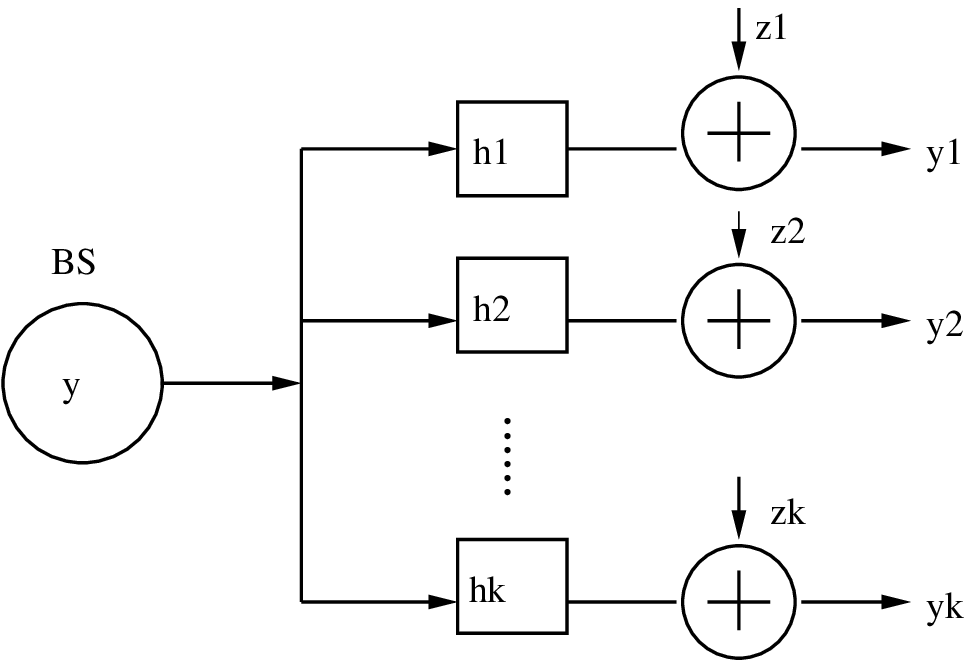}
        \centerline{BC, $\bz_i\sim
        \mathcal{N}(0,\sigma^2\bI_{N_r})$,}\medskip\\
        \centerline{$q_{t} \sum_{i=1}^{K}\bh_{o}^{\dag}\bQ_i^{\text{b}}\bh_{o}+q_{u}\sum_{i=1}^{K}\text{tr}(\bQ_i^{\text{b}}) \leq P$}\medskip
    \end{minipage}
    \hfill
    \begin{minipage}[b]{0.48\linewidth}
        \centering
        \psfrag{h1}{$\bH_{1}^{\dag}$}
        \psfrag{h2}{$\bH_{2}^{\dag}$}
        \psfrag{hk}{$\bH_{K}^{\dag}$}
        \psfrag{u1}{$\bx_{1}$}
        \psfrag{u2}{$\bx_{2}$}
        \psfrag{uk}{$\bx_{K}$}
        \psfrag{Z}{$\bz$}
        \psfrag{y}{$\by$}
        \psfrag{BS}{$\text{BS}$}
        \vspace{8mm}
        \includegraphics[width = 80mm,height=52mm]{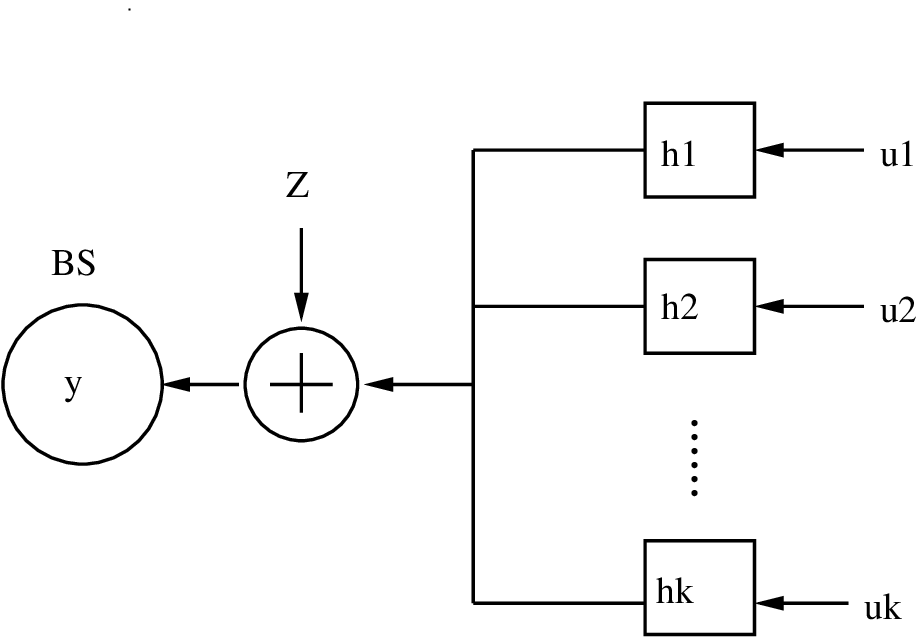}
        \centerline{Dual MAC, $\bz\sim
        \mathcal{N}(0,q_{t}\bR_o+q_{u}\bI_{N_t})$,}\medskip\\
        \centerline{$\sum_{i=1}^{K}\text{tr}(\bQ_i^{\text{m}})\sigma^2\leq P,~\bR_o=\bh_o\bh_o^H$}\medskip
    \end{minipage}
     \caption{The system models for Problem \ref{Prob:fixq} and Problem \ref{eq:objfunMACdual}, where $q_{t}$ and $q_{u}$ are constant, and $\bR_o=\bh_{o}\bh_o^{\dag}$.}
     \label{fig:sysmodelMAC}
\end{figure}

\begin{figure}
      \centering
      \psfrag{q1}{$q^{(1)}_{t},q^{(1)}_{u}$}
      \psfrag{q2}{$q^{(2)}_{t},q^{(2)}_{u}$}
      \psfrag{qi}{$q^{(n)}_{t},q^{(n)}_{u}$}
      \psfrag{q3}{$q^{(3)}_{t},q^{(3)}_{u}$}
      \psfrag{Qm1}{$\bQ_{i,(1)}^{\text{m}}$}
      \psfrag{Qm2}{$\bQ_{i,(2)}^{\text{m}}$}
      \psfrag{Qm3}{$\bQ_{i,(3)}^{\text{m}}$}
      \psfrag{Qmn}{$\bQ_{i,(n)}^{\text{m}}$}
      \psfrag{Qb1}{$\bQ_{i,(1)}^{\text{b}}$}
      \psfrag{Qb2}{$\bQ_{i,(2)}^{\text{b}}$}
      \psfrag{Qbn}{$\bQ_{i,(n)}^{\text{b}}$}
      \psfrag{......}{$\cdots$}
     \includegraphics[width = 85mm,height=50mm]{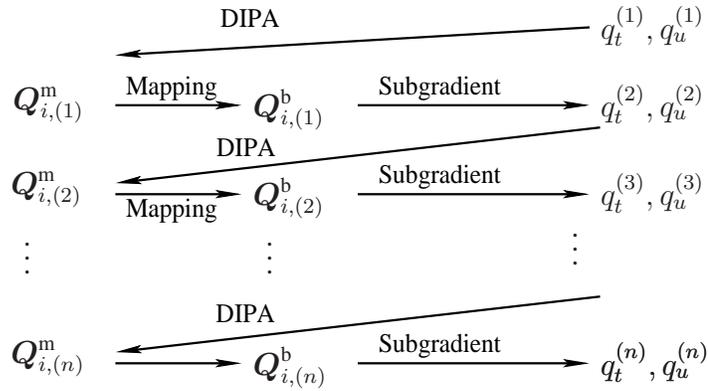}
     \caption{The flow chart for the SIPA algorithm, where $\bQ_{i,(n)}^{\text{b}}$ and $\bQ_{i,(n)}^{\text{n}}$ denote the
     transmit signal covariance matrices of SU$_i$ for the BC and MAC at the $n$th step, respectively.}
     \label{fig:SIPAchart}
\end{figure}

\begin{figure}
     \centering
     \includegraphics[width = 120mm]{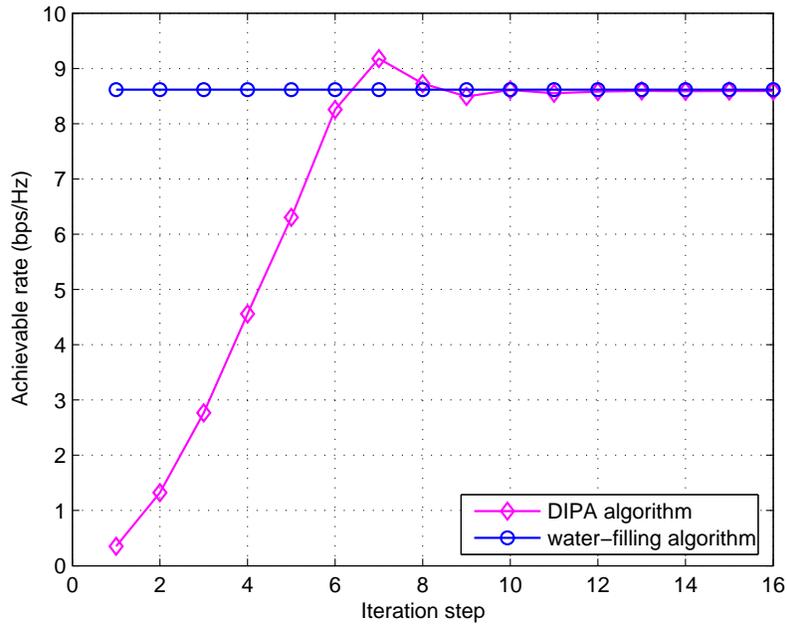}
     \caption{Comparison of the optimal achievable rates obtained by the DIPA and the water-filling algorithm in a MIMO channel ($N_t=N_r=4$, $K=1$ and $P_u$=10 dB).}
     \label{fig:compwf}
\end{figure}

\begin{figure}
     \centering
     \includegraphics[width = 120mm]{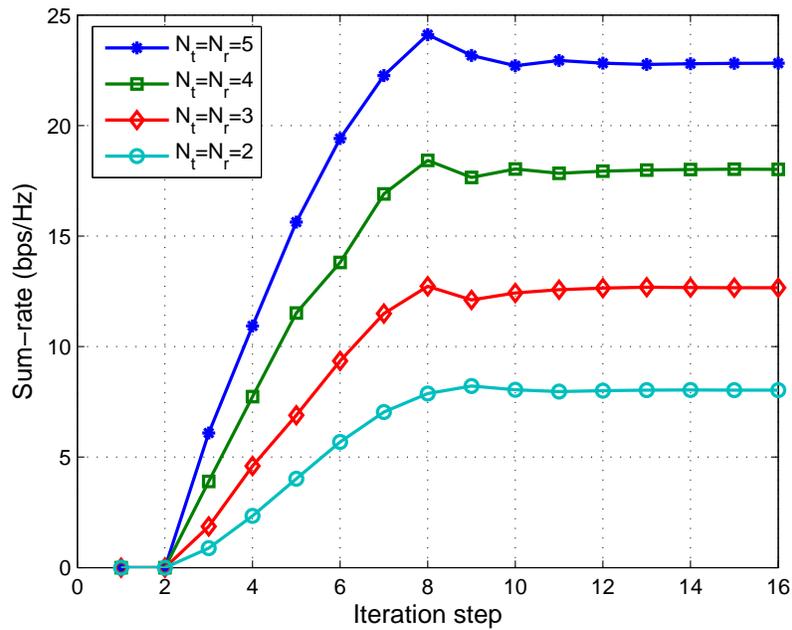}
     \caption{Convergence behavior of the DIPA algorithm ($K=20$ and $P_u=10$ dB).}
     \label{fig:dipaconv}
\end{figure}

\begin{figure}
     \centering
     \includegraphics[width =120mm]{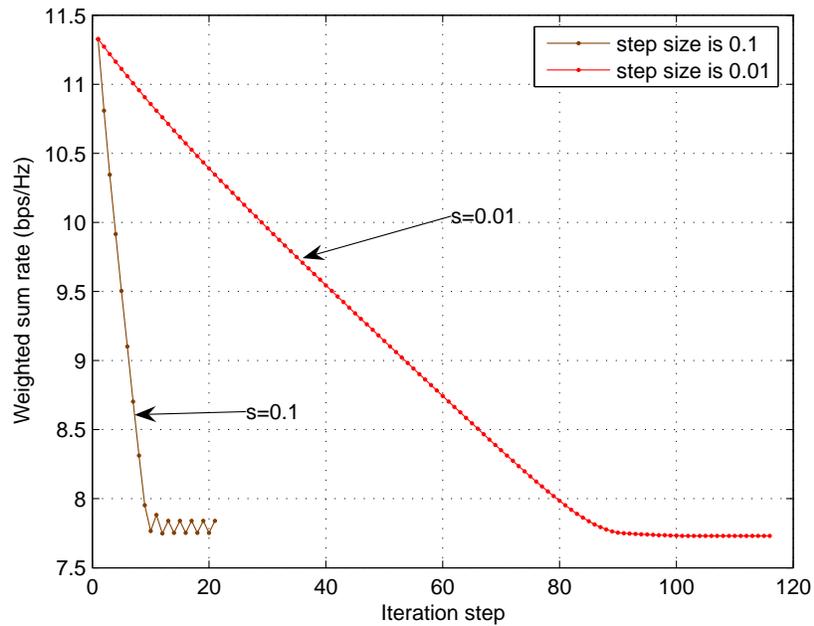}
     \caption{Convergence behavior of the SIPA algorithm ($N_t=5$, $K=5$, $N_r=3$, $w_1=5$, and $w_i=1$, for $i\neq 1$).}
     \label{fig:rate}
\end{figure}

\begin{figure}
     \centering
     \includegraphics[width = 120mm]{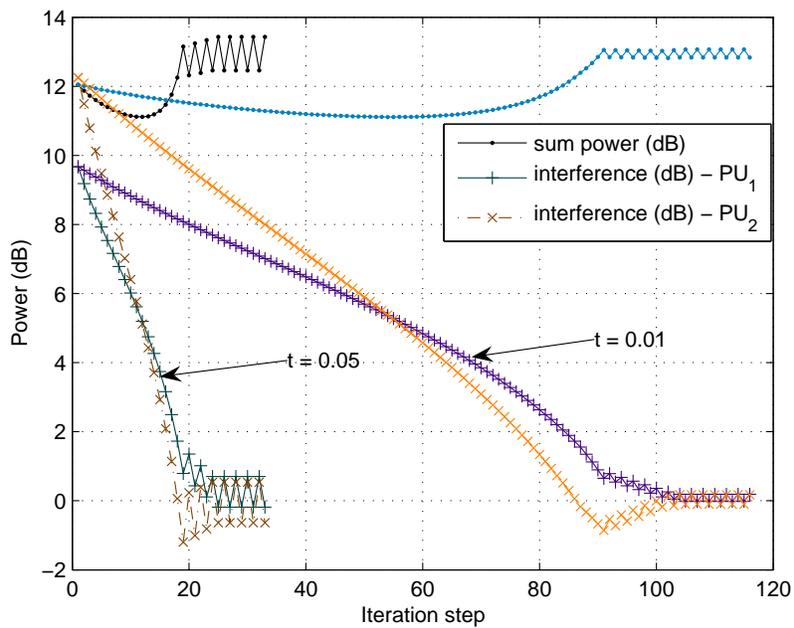}
     \caption{The convergence behavior of the sum power at the BS and the interference at the PU for the SIPA algorithm ($N_t=5$, $K=5$, $N_r=3$, $w_1=5$, and $w_i=1$ with $i\neq 1$).}
     \label{fig:power}
\end{figure}

\begin{figure}
     \centering
     \includegraphics[width = 120mm]{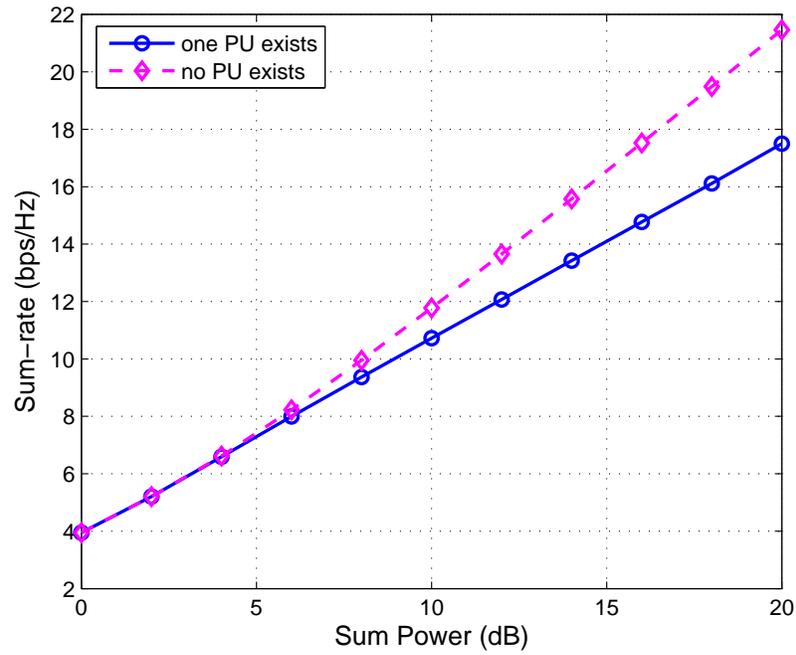}
     \caption{Achievable sum rates versus sum power in the single PU case and the case with no PU ($N_t=5$, $K=5$, $N_r=3$).}
     \label{fig:sumpowerincr}
\end{figure}

\begin{figure}
     \centering
     \includegraphics[width = 120mm]{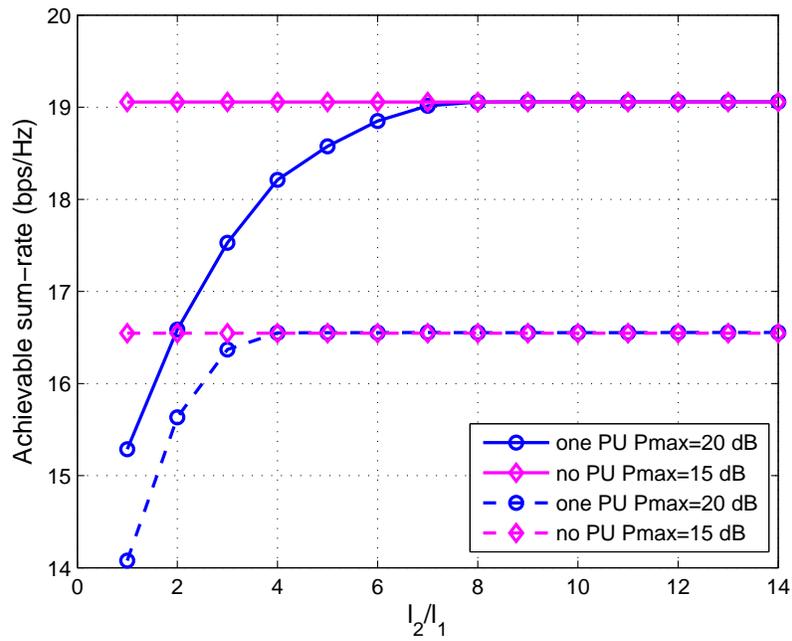}
     \caption{Achievable sum rates versus the ratio of $l_2/l_1$ using the SIPA algorithm ($N_t=5$, $K=5$, $N_r=3$).}
     \label{fig:dist}
\end{figure}

\end{document}